\documentclass{article}

\usepackage[normalem]{ulem}
\usepackage{mathptmx} 
\usepackage{geometry}
\usepackage{fancyhdr}
\usepackage{amsmath,amsfonts}
\usepackage{url}
\usepackage{mathptmx} 

\usepackage{graphicx}
\usepackage{textcomp}
\usepackage{booktabs}
\usepackage{color,soul}
\usepackage{xcolor}

\usepackage{caption}

\usepackage{commath}
\usepackage{comment}
\usepackage{paralist} 
\usepackage{xcolor}
\usepackage{listings}
\usepackage[linesnumbered,ruled,vlined]{algorithm2e}
\usepackage{caption}
\usepackage{ifthen}
\usepackage{subcaption}

\usepackage{float}
\usepackage{hyperref}

\definecolor{mGreen}{rgb}{0,0.6,0}
\definecolor{mGray}{rgb}{0.5,0.5,0.5}
\definecolor{mPurple}{rgb}{0.58,0,0.82}
\definecolor{backgroundColour}{rgb}{0.95,0.95,0.92}

\lstdefinestyle{CStyle}{
    backgroundcolor=\color{backgroundColour},   
    commentstyle=\color{mGreen},
    keywordstyle=\color{magenta},
    numberstyle=\tiny\color{mGray},
    stringstyle=\color{mPurple},
    basicstyle=\footnotesize,
    breakatwhitespace=false,         
    breaklines=true,                 
    captionpos=b,                    
    keepspaces=true,                 
    numbers=left,                    
    numbersep=5pt,                  
    showspaces=false,                
    showstringspaces=false,
    showtabs=false,                  
    tabsize=2,
    language=C
}
\newcommand{\allowcomments}{0} 

\newcommand{\scheme}{\textsc{AOK}}
\newcommand{\cpumodel}{Intel Xeon Phi 7290}
\newcommand{\opensslver}{1.1.0f}

\ifthenelse{\equal{\allowcomments}{0}}{
\newcommand{\fm}[1]{\textcolor{blue}{}}
\newcommand{\am}[1]{\textcolor{red}{}}
\newcommand{\cc}[1]{\textcolor{green}{}}
\newcommand{\sk}[1]{\textcolor{cyan}{}}
\newcommand{\ej}[1]{\textcolor{pink}{}}
\newcommand{\hs}[1]{\textcolor{olive}{}}
}{
\newcommand{\fm}[1]{\textcolor{blue}{FM: #1}}
\newcommand{\am}[1]{\textcolor{red}{AM: #1}}
\newcommand{\cc}[1]{\textcolor{green}{CC: #1}}
\newcommand{\sk}[1]{\textcolor{cyan}{SK: #1}}
\newcommand{\ej}[1]{\textcolor{pink}{EJ: #1}}
\newcommand{\hs}[1]{\textcolor{olive}{HS: #1}}
}

\author{Farabi Mahmud, Sungkeun Kim, Harpreet Singh Chawla, 
	\\Chia-Che Tsai, EJ Kim, Abdullah Muzahid
	\\\textit{Texas A\&M University,}
	\\\href{mailto:farabi@tamu.edu}{farabi@tamu.edu}}

\title{Attack of the Knights:\\A Non Uniform Cache Side-Channel Attack}

\begin{document}

\maketitle
\begin{abstract}

For a distributed last-level cache (LLC) in a large multicore chip, the access time to one LLC bank can significantly differ from that to another due to the difference in physical distance. In this paper, we successfully demonstrated a new distance-based side-channel attack by timing the AES decryption operation and extracting part of an AES secret key on an Intel Knights Landing CPU. 
We introduce several techniques to overcome the challenges of the attack, including the use of multiple attack threads to ensure LLC hits, to detect vulnerable memory locations, and to obtain fine-grained timing of the victim operations. 
While operating as a covert channel, this attack can reach a bandwidth of 205 kbps with an error rate of only 0.02\%. We also observed that the side-channel attack can extract 4 bytes of an AES key with 100\% accuracy with only 4000 trial rounds of encryption.
\end{abstract}

\section{Introduction}

\begin{figure}[t]
    \vspace{10pt}
    \centering
    \includegraphics[width=0.9\columnwidth]{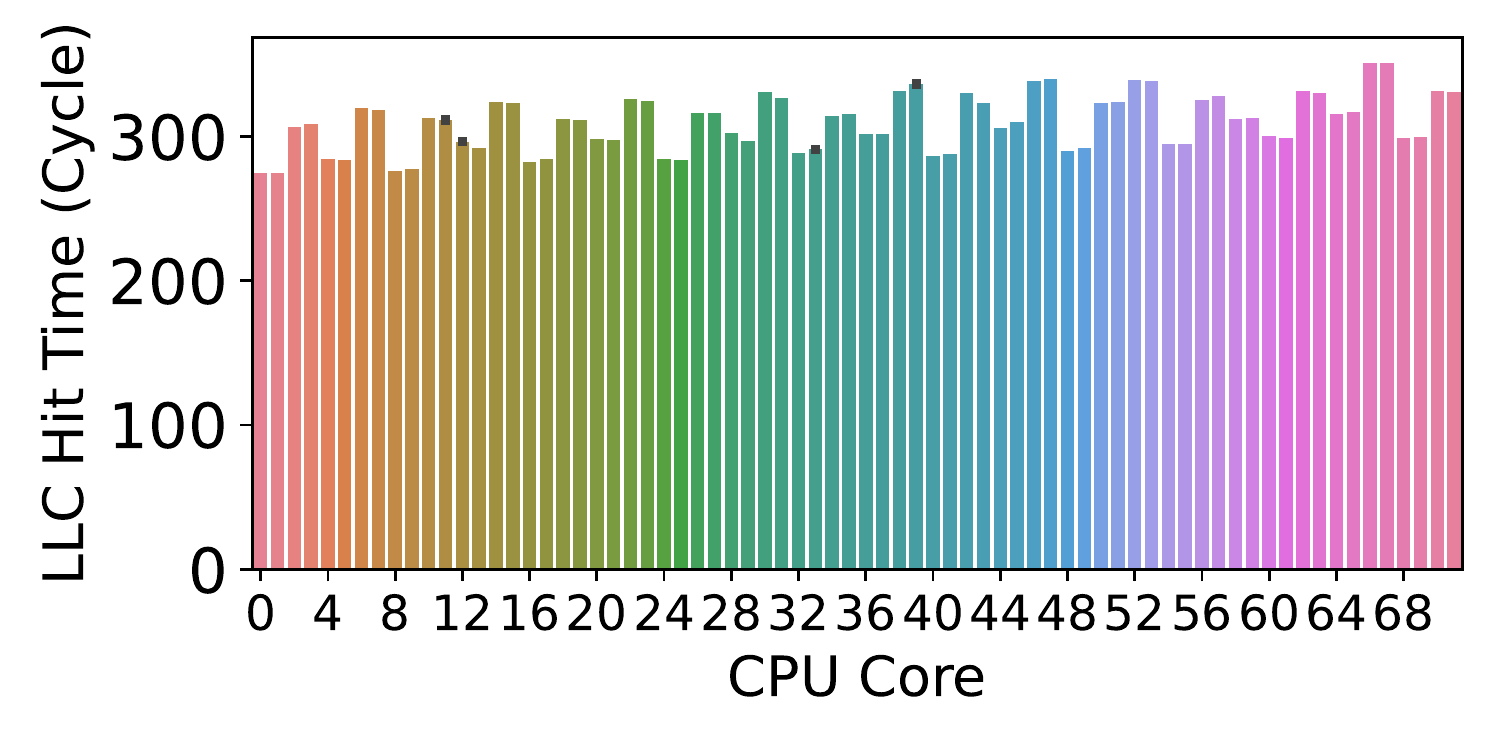}
    \caption{LLC hit time measured in cycles when accessing the same physical address ({\tt 0x1000000000}) from different cores (core 0 to 71) in the \cpumodel{} CPU. The latency numbers are averaged over 10,000 samples.} 
    \label{fig:all-core-one-addr}
\end{figure}

The ever-shrinking feature size in CMOS process technology has enabled the integration of a large number of cores and caches in a single chip~\cite{borkar1999design,hoeneisen1972fundamental}.
Large-scale multicores are equipped with a large last-level cache (LLC) to alleviate expensive off-chip accesses. 
As an example, AMD Ryzen comes with 256 MB LLC whereas Intel Xeon Phi 7200 series has 36 MB LLC.~\cite{amdryze,xeonphi} 
Such an LLC is distributed over multiple banks connected through a network-on-chip (NoC) to reduce access latency and improve core isolation, 
referred to as a Non-Uniform Cache Access (NUCA) architecture.
With a distributed LLC, memory-level parallelism is improved by allowing concurrent requests to different banks.
However, on the negative side, access latency to different banks may vary depending on the distance from the requesting core as seen in Figure~\ref{fig:all-core-one-addr}. 
In this paper, we set out to investigate whether this latency difference can lead to security vulnerabilities in large-scale multicore machines.


When it comes to security vulnerabilities related to caches, side-channel attacks are the most notorious ones. Numerous 
cache side-channel attacks~\cite{liu2015last,schwarz2017malware,brasser2017software,kayaalp2016high,aciiccmez2008vulnerability,tromer2010efficient,osvik2006cache,neve2006advances,percival2005cache} were discovered and shown to be significant threats to data security, especially during the past few years.
The most common form of cache side-channel attacks involves timing the cache access latency which depends on the state of the target cache lines. Take Flush+Reload~\cite{gruss2015cache,zhang2014cross,yarom2014recovering,van2015just,irazoqui2014wait,yarom2014flush+,gullasch2011cache} for example. The attacker tries to observe the target shared lines being accessed by the victim program, by detecting whether reloading the line incurs a cache hit or miss. Most existing cache-based side-channel attacks exploit the timing difference caused by cache states. Numerous defense techniques~\cite{kiriansky2018dawg,liu2016catalyst,cloakhtm,cat,catxeon,pageplacement, caesar} have been proposed to eliminate such timing differences.
Furthermore, most of the existing cache side-channel attacks have not been demonstrated or explored in Non-Uniform Cache Access (NUCA) architecture.
Recently, Dai et al.~\cite{dont-mess-around} 
showed how network contention in NUCA architecture can be utilized to create timing differences and subsequently, a covert and side-channel. 
As we see more and more large-scale processor chips~\cite{skylake,cascadelake,nurion,ampere-altra-review,cori,xeonphi,skylake-sp}, attackers will turn their attention to similar attacks which have the potential to circumvent prior defenses. However, as these multicore chips scale up, contention-based NUCA attacks will become more difficult due to noises in the on-chip communication. 
In this paper, we showed that under such circumstances, there is still a possibility of a new side-channel attack in large-scale multicore chips that rely on physical distances in cache banks as opposed to network contention. 



In this paper, we demonstrated a novel distance-based side-channel attack in large-scale NUCA architecture that may exist even when other cache-based side-channel attacks have been rendered ineffective.
As a simplified case, let us consider Algorithm~\ref{alg:psuedoattack}. It shows the pseudo-code of a side-channel attack that relies on physical distance in a NUCA machine. 
In this example, an address is determined by the victim according to a specific bit of a secret.
The address can be mapped to the LLC bank of the nearest core when the bit is 0, or the farthest core when the bit is 1. 
Therefore, by timing the access latency, an attacker can infer the secret bit. 

To further illustrate this scenario, Figure~\ref{fig:all-core-one-addr} shows access latency to the same address from different cores.
We collected these latency numbers from the \cpumodel{} CPU.
This CPU model belongs to Intel's Knights Landings line and has a multicore architecture with at least 64 cores (the CPU we tested has 72 cores). 
The figure shows that the LLC hit latency for the same address has a range between 280--350 cycles, and this pattern is generally stable across cores.
If an attacker can measure the access latency in the victim code (as shown in Algorithm~\ref{alg:psuedoattack}), he/she will be able to guess the secret bit by telling whether the addresses fall into a near or far LLC bank. We refer to this as NUCA distance-based side-channel.

We demonstrate this side-channel in \cpumodel{} using AES code. 
We have addressed two major technical challenges for this attack, namely (i) the overlapping of memory accesses to LLC banks, and (ii) the difficulty of timing AES operations.
To address these challenges, we employed different techniques.
We have used AdaBoost~\cite{adaboost} model to differentiate between favorable and unfavorable cases in the presence of overlapping LLC accesses. With multiple samples and majority voting, we get 100\% confidence in predicting the vulnerable access pattern as described in \S\ref{sec:accuracy-ml-classifier}. 
To efficiently time only the region of interest within the decryption function, we have utilized multiple attack threads and time part of the AES decryption operation.
Our proof-of-concept (PoC) attack code can accurately extract 4 bytes of any AES key with 4,000 decryption trials using a sequence of random plaintexts.
We disclosed this vulnerability to Intel Product Incident Response Team (PSIRT). The team subsequently confirmed this vulnerability.

It should be noted that our attack is different from prior attacks. For example, in the case of Prime+Probe and Dai et al., because the attacks depend on sharing resources, the attack can be mitigated by partitioning or clustering the resources. 
Dai et al. can also be mitigated through software defenses by exploiting specific scheduling patterns in NoC routers. 
However, NUCA distance-based side-channel attacks cannot be completely mitigated by partitioning or clustering as long as the attacker is able to time the victim. 
The success of the attack depends on the capability of the attacker to time the victim, not the sharing of resources. 

For responsible disclosure, we have revealed the attack details and provided the PoC attack code to Intel and received acknowledgment from Intel.

\begin{algorithm}[tbp]
\footnotesize
\KwIn{BitMask}
 \KwData{Secret}
 \eIf{(Secret \& BitMask) == 1}{  
    $Addr$ = ${Addr}_{Near}$ mapped to the LLC bank of the nearest core\;
 }{
    $Addr$ = ${Addr}_{Far}$ mapped to the LLC bank of the farthest core\;
 }
 Load($Addr$)\;
\caption{Pseudo code of a victim function which is vulnerable to a side-channel based on the difference of LLC access time due to distance.}
\label{alg:psuedoattack}

\end{algorithm}

In summary, we make the following contributions:
\begin{compactitem}
    \item We demonstrated 
    a {\em new} distance-based side-channel attack on NUCA architecture on an Intel Knights Landing CPU against a vulnerable AES implementation. The proof-of-the-concept code can accurately extract the 4 bytes of the AES key with only 4,000 decryption trials using a sequence of random plaintexts.

    \item
    We identified and addressed two technical challenges to performing a robust NUCA distance-based side-channel attack.

    \item
    We developed several techniques including the use of a machine learning model to classify latency that consists of multiple cache accesses, use of a separate timing thread for fine-grained timing of the victim operations, and use of a remote thread to force L1D misses but LLC hits for the target cachelines. 

\end{compactitem}

The rest of the paper is organized as follows: \S\ref{sec:background} explains The necessary background of NUCA architecture, different cache side-channel attacks, AES algorithm; \S\ref{sec:target-architecture} explains the details of the target architecture;
\S\ref{sec:attack-assumptions} explains the  assumptions for the attack model; \S\ref{sec:attack} gives detailed examples of attacks in a microarchitectural simulator and a real machine and shows other vulnerable algorithms;
\S\ref{sec:experiments} shows different experiments to prove our claims;
Finally, we discuss some possible defenses in \S\ref{sec:possible-defense}. 

\section{Background \& Related Works}
\label{sec:background}

In this section, we describe the background of non-uniform cache access (NUCA) architecture, cache-based side-channel attacks, and NoC-based side-channel attacks.

\subsection{Non-Uniform Cache Access Architecture}

The Non-Uniform Cache Access (NUCA) Architecture is designed to reduce the memory access time by increasing the cache capacity and thus increasing the cache hit rate~\cite{jaleel2006last,muralimanohar2007interconnect,amdryze}.
However, as the bandwidth and the number of ports are limited for individual physical cache banks, multicore processor chips adopt the design of having physically separated banks for shared last-level caches (LLC), which are interconnected with network-on-chip (NoC).
As a result, cache access latency to different cache banks in a NUCA architecture shows disparity~\cite{kim2002adaptive}, as the memory operations involve network protocols and messaging within NoC.  

\begin{figure}[t]
  \subfloat[Intel Skylake Die.]{
	   \centering
	   \includegraphics[width=0.45\columnwidth]{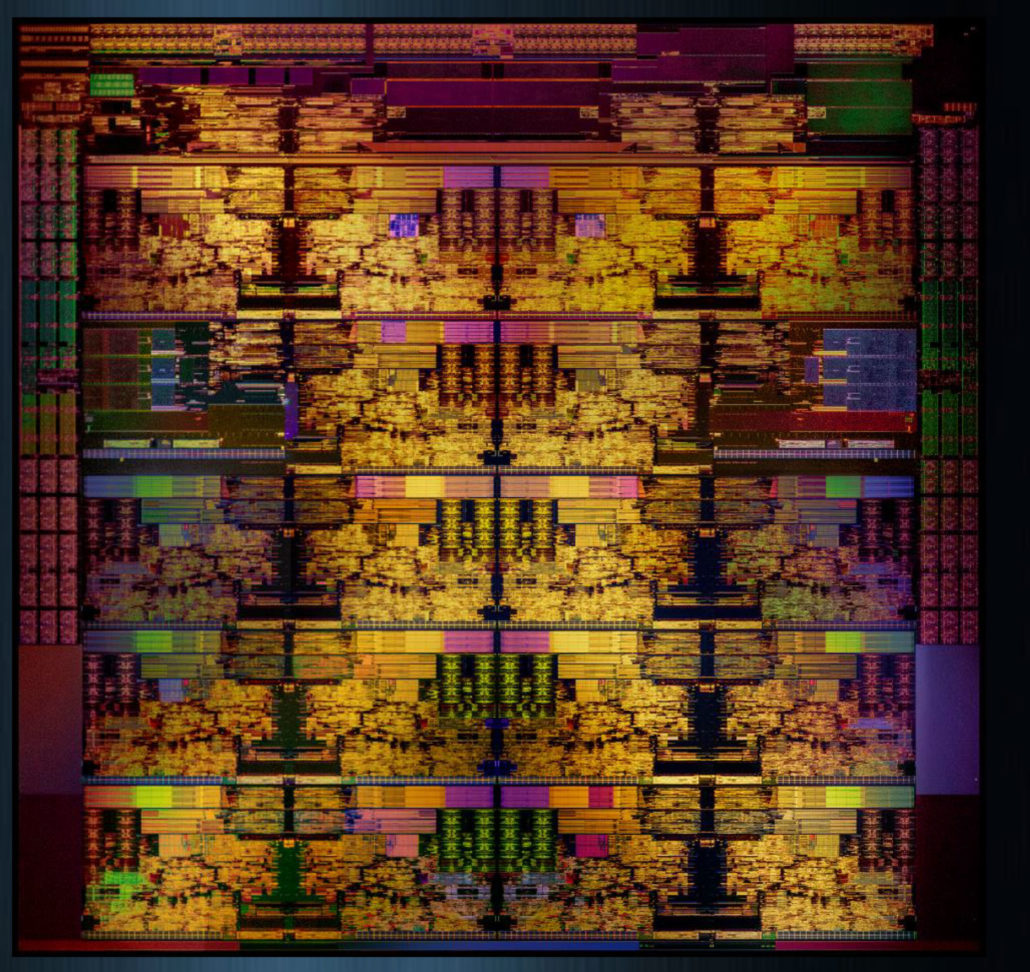}
   }
  \subfloat[Intel Skylake Mesh Network.]{
	   \centering
	   \includegraphics[width=0.45\columnwidth]{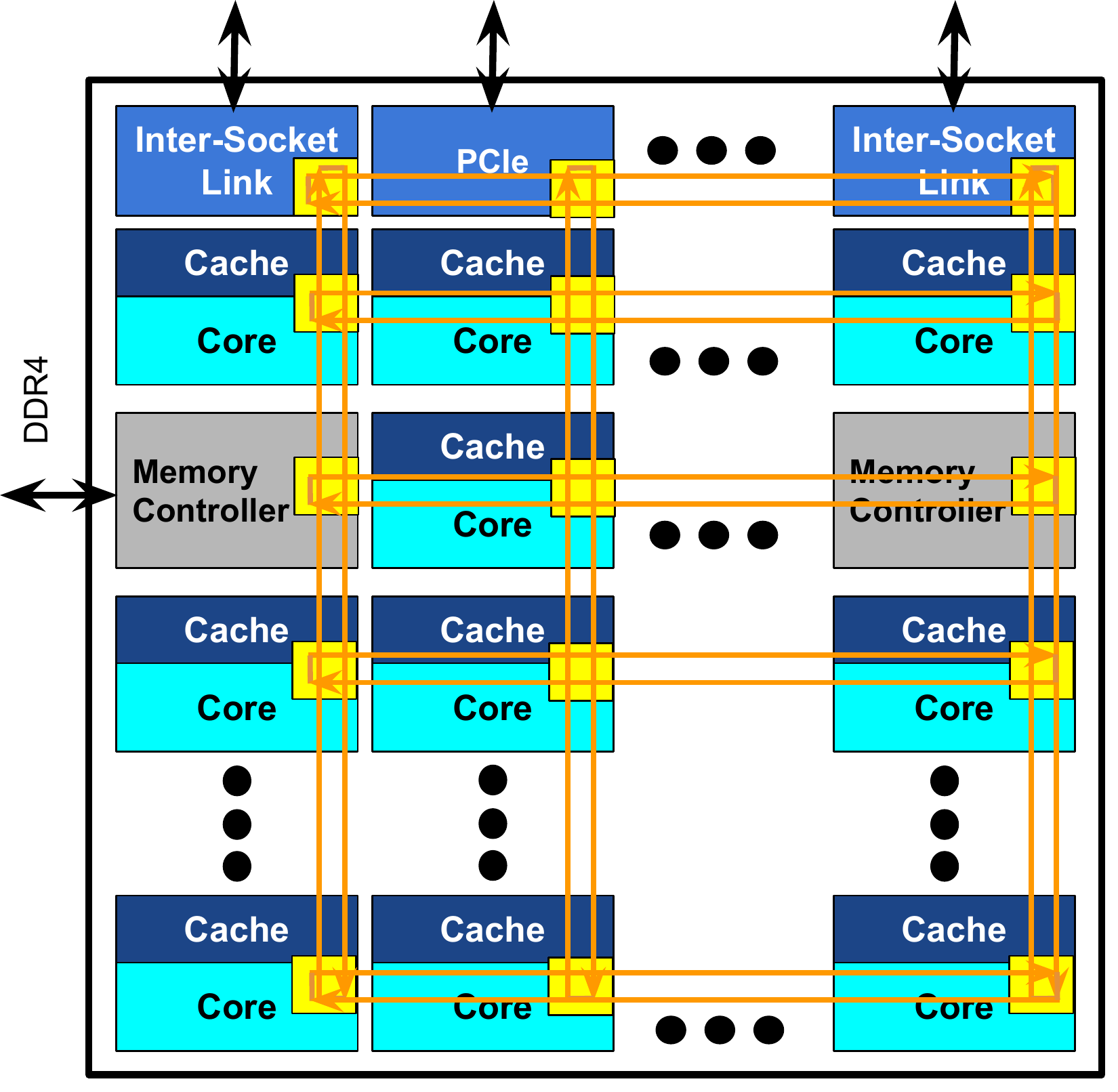}
   }
\caption{NUCA Architecture in Intel's Skylake system. In the architectural diagram (b), each cache block consists of private L1, L2 and a shared L3/LLC
bank.}
	\label{fig:skylake}
\end{figure}

In this paper, we particularly focus on the Mesh-based interconnect topology of the Skylake-SP and Skylake-X processors of Intel, which have 28 cores~\cite{skylake} as shown in Figure~\ref{fig:skylake}, and AMD's 32-core EPYC processors~\cite{amdnuca}.
These processors use a different NoC design as the traditional ring-based architecture of Intel processors, and thus exhibit completely different access latency patterns from prior Intel processors. 
\fm{according to dont-mess-around paper, mesh is also implemented with bidirectional ring}



\subsection{Cache Side-channel Attacks}

Existing cache side-channel attacks exploit the fact that the internal states of CPU caches, including TLB occupancy, cache eviction, cache replacement states, and memory controller buffers can be observed in the microarchitecture~\cite{tlb-attack,osvik2006cache,lru-sidechannel,coherence-protocol}.
The attackers that use these side channels often time the latency of memory operations after the victims have accessed the TLB or the cache, and then use the information to infer the behaviors of the victims or even the secrets inside the victim programs.

For a side channel, the sender (or the ``victim'') and the receiver (or the ``attacker'') use the observable states of a system to encode and decode the information that they intend to communicate.
Such communication typically requires a certain level of synchronization, especially if the sender and the receiver are communicating more than one bit.
Synchronization is required between the sender sending the information by affecting the states of the system, and the receiver receiving the information by detecting the state changes and decoding the information.
Between each bit being transmitted with the side channel, the participants also need to reset/refresh the state of the system because the receiver cannot distinguish the state changes caused by transmitting multiple bits from the sender.


Below, we explain some of the existing cache side-channel attack techniques, and how they transmit information:

\begin{compactitem}


\item
\textbf{Prime+Probe
~\cite{liu2015last,schwarz2017malware,schwarz2017malware,brasser2017software,kayaalp2016high,aciiccmez2008vulnerability,tromer2010efficient,osvik2006cache,neve2006advances,percival2005cache}:
} 
This attack exploits the observable eviction in the shared cache when CPU needs to replace one of the attacker's cache lines with the cache line of the victim.
To ensure the observable eviction, the attacker needs to first \emph{prime} part of or the entirety of the cache, so that cache replacement will occur on one of the attacker's cache lines.


\item
\textbf{Flush+Reload~
\cite{gruss2015cache,zhang2014cross,yarom2014recovering,van2015just,irazoqui2014wait,yarom2014flush+,gullasch2011cache}:
}
This attack depends on the attacker and the victim sharing the same data in the memory, and thus when the CPU brings the data into the cache for the victim, the attacker can access the same cache line afterward.
Given that the cache line may be previously brought in for the attacker, the attacker will use instructions like CLFLUSH to
remove the cache line prior to the potential operations in the victim.
The sharing of data between the attacker and the victim can be a result of shared libraries or physical memory deduplication.


\item
\textbf{Evict+Reload:} 
Similar to Flush+Reload, Evict+Reload depends on the attacker and the victim sharing the same cache lines.
The only difference is that the attacker first performs multiple memory accesses to evict the target cache line, instead of flushing the cache line using CLFLUSH. This technique is useful when CLFLUSH is not available on the specific architecture.



\item
\textbf{Flush+Flush~\cite{gruss2016flush+}}:
Flush+Flush also depends on the attacker and the victim sharing the cache line within the CPU, yet it exploits the difference of access latency due to the cache coherence protocol. 
In Flush+Flush, the attacker times the CLFLUSH instruction, aside from using it as a measure of resetting the cache states. 
If the target cache line was never accessed by a remote CPU core and thus was never brought into another private cache, the access latency will be much lower than that of the scenario in which the cache line is held in another private cache.


\end{compactitem}

The mitigations of these side-channel attacks are mostly based on two approaches.
The first approach prevents the attackers from observing the information decoded inside the timing of cache operations, such as adding extra latency to cache access
\cite{godfrey2013server,zhang2011predictive}.
The second approach prevents the attacker and the victim from sharing resources such as the last-level cache and thus eliminates the medium which they can use for communication
\cite{liu2016catalyst,kiriansky2018dawg,wang2007new, kim2012stealthmem}. 

\subsection{Other NoC-based Side-Channel Attacks} \label{sec:nocbased_sidechannel}

In NUCA architecture, NoC has been used for attackers the detect the access latency chance after the eviction or fetching of cache lines, similar to Prime+Probe~\cite{reinbrecht2016side}.
In order to mitigate such a side channel,
prior work~\cite{reinbrecht2016side} proposes obfuscating the NoC access patterns by swapping routing algorithms when the attack is suspected.


Another type of explored side channels in NoC
is the exploitation of NoC and cache contention, especially in a Ring interconnect topology~\cite{paccagnella2021lotr}.
These side channels exploit the interference of NoC traffics,
and thus can be mitigated by isolating the network traffics of different processes or CPU cores~\cite{wassel2013surfnoc}.

\subsection{Comparison with Existing Attacks} \label{sec:comp_existing_attack}

To show the difference between our NUCA distance-based side-channel attack and the existing cache side-channel attack (we use Prime+Probe~\cite{percival2005cache} as an example) and NoC-based attack described (Dai et al.~\cite{dont-mess-around} as an example), we compare these attacks according to their attack models, attack assumptions, and the possible defenses.



\textbf{Attack Models:}
The main difference between these attacks is in the way the sender (the ``victim'') transmits information to the receiver (the ``attacker''). 
The cache side-channel attacks such as Prime+Probe primarily transmit information by evicting or populating specific cache lines.
For the NOC attack by Dai et al., the information is transmitted through the interference of traffic on the shared NoC rings inside a ring-based Mesh interconnect.
On the other hand, our attack does not transmit information through changes of cache states or interference, but rather through directly timing the memory access operations inside the victim to infer the secret from the difference in cache latency.
This difference is fundament to how this attack is performed as well as the possible defense to the attack.


\textbf{Attack Assumptions}:
For the existing cache side channels (e.g., Prime+Probe), the attacker and the victim needs to share certain resources and as a result, causing the victim's behavior to be observable by the attacker.
The NoC attack by Dai et al. requires the attacker to interfere with the traffic of the victim.
Our NUCA attack does not require any contention or interference, but does require the attacker to be able to time the victim operations directly.
In addition, we show that, end-to-end latency of victim operations can be too coarse-grained for the attacker to infer the victim's secret accurately, so the attack will require fine-grained timing information about the latency of individual memory access or at least groups of memory access.
This is one of the important hurdles that the attacker needs to clear in order to exploit this side channel.


\textbf{Possible Defenses:}
The existing cache side channels are typically mitigated by either un-sharing the resources or obfuscating the timing in the CPU cache.
For the NoC attack by Dai et al., interference of NoC traffic can be mitigated by changing the scheduling policies within the NoC or isolating the NoC channels between the attacker and the victim.
For our NUCA attack, the only effective way is to block the attacker from observing any meaningful patterns in the timing of the victim operations.
Along the same line of thinking, oblivious RAM (ORAM)~\cite{oram} can be used to remove the access patterns completely, and thus can be used to mitigate all three attacks simultaneously.

\subsection{Advanced Encryption Standard (AES)}
Advanced Encryption Standard (AES) is a block cipher that is widely used. This was initially proposed in 1997 and later adopted as the standard by the National Institute of Standards and Testing~\cite{aes-original}. 
Some ISAs started including modified instructions in their ISAs to accommodate AES operations~\cite{aes-isa}.

AES can have 128-bit, 192-bit, or 256-bit keys and can comprise of 10 rounds, 12 rounds, or 14 rounds of encryption/decryption respectively. 
Each round can be broken into different steps. These include - 
\begin{enumerate}
    \item KeyExpansion - Different keys used in each round are calculated from the original cipher key using the predetermined AES Key Schedule 
    \item AddRoundKey - During this step, Round Key is XOR'ed with the current state
    \item SubBytes - A lookup table is used to nonlinearly substitute each byte with some specific values
    \item ShiftRows - Last three rows of the state are shifted cyclically 
    \item MixColumns - Mixing operation that combines four bytes in each column 
\end{enumerate}
In the first round KeyExpansion and AddRoundKey are done. 
In next consecutive 9, 11, or 13 rounds (based on the number of bits in the cipher key), SubByte, ShiftRow, MixColumn and AddRoundKey steps are done. 
In the final round, only SubByte, ShiftRow and AddRoundKey operation is executed. In our attack, we target the last round of decryption. 

\section{Target Architecture Details}
\label{sec:target-architecture}
In this section, we explain the architectural details of the target architecture that we have used to implement the attack. We also discuss how this makes other architectures vulnerable as well. 
The ~\cpumodel{} is based on the Intel Knights Landing family that was launched in 2016. 
Many HPC servers and cloud providers are still using Knights Landing processors in high performance computing including Cori~\cite{cori} from National Energy Research Scientific Computing Center, Stampede-2~\cite{stampede2} from the University of Texas, Trinity~\cite{trinity} from Oak Ridge National Laboratory or Nurion~\cite{nurion} from Korea Institute of Science \& Technology Information. 
However, the principle that is exploited in this attack, the latency difference due to mesh on chip network, is carried into current and future generations of Intel Skylake-SP~\cite{skylake}, Cascade Lake~\cite{cascadelake} and even newer processors. 
\subsection{Memory Configuration}

\subsubsection{MCDRAM Configuration}
To accommodate computationally intensive programs, Intel Knights Landing series processor started including high bandwidth on-package memory in the form of Multi Channel DRAM (MCDRAM). 
MCDRAM is a high bandwidth memory that supports multiple channels at the same time. 
Even though this memory is physically located within the same package, it is situated on a separate piece of silicon die.\ej{check please if it is correct}
In ~\cpumodel~ we have 16Gb of MCDRAM on-package.
We have three different memory modes depending on the usage of the MCDRAM. These three modes are - 
\begin{enumerate}
    \item \textbf{Cache Mode} - MCDRAM used only as shared Cache memory 100\% cache, 0\% memory
    \item \textbf{Flat Mode} - MCDRAM used as 100\% memory and 0\% cache. No separate DDR memory is required.
    \item \textbf{Hybrid Mode} - MCDRAM is used as 25\%/50\% cache and the rest is used as memory. 
\end{enumerate}
In our experiments, we have used the MCDRAM in flat modes, however, this can be extended to any memory mode. 
We leave the exploration of our attack in different memory modes for future works. 

\begin{figure}[ht]
    \centering
    \includegraphics[width=\columnwidth]{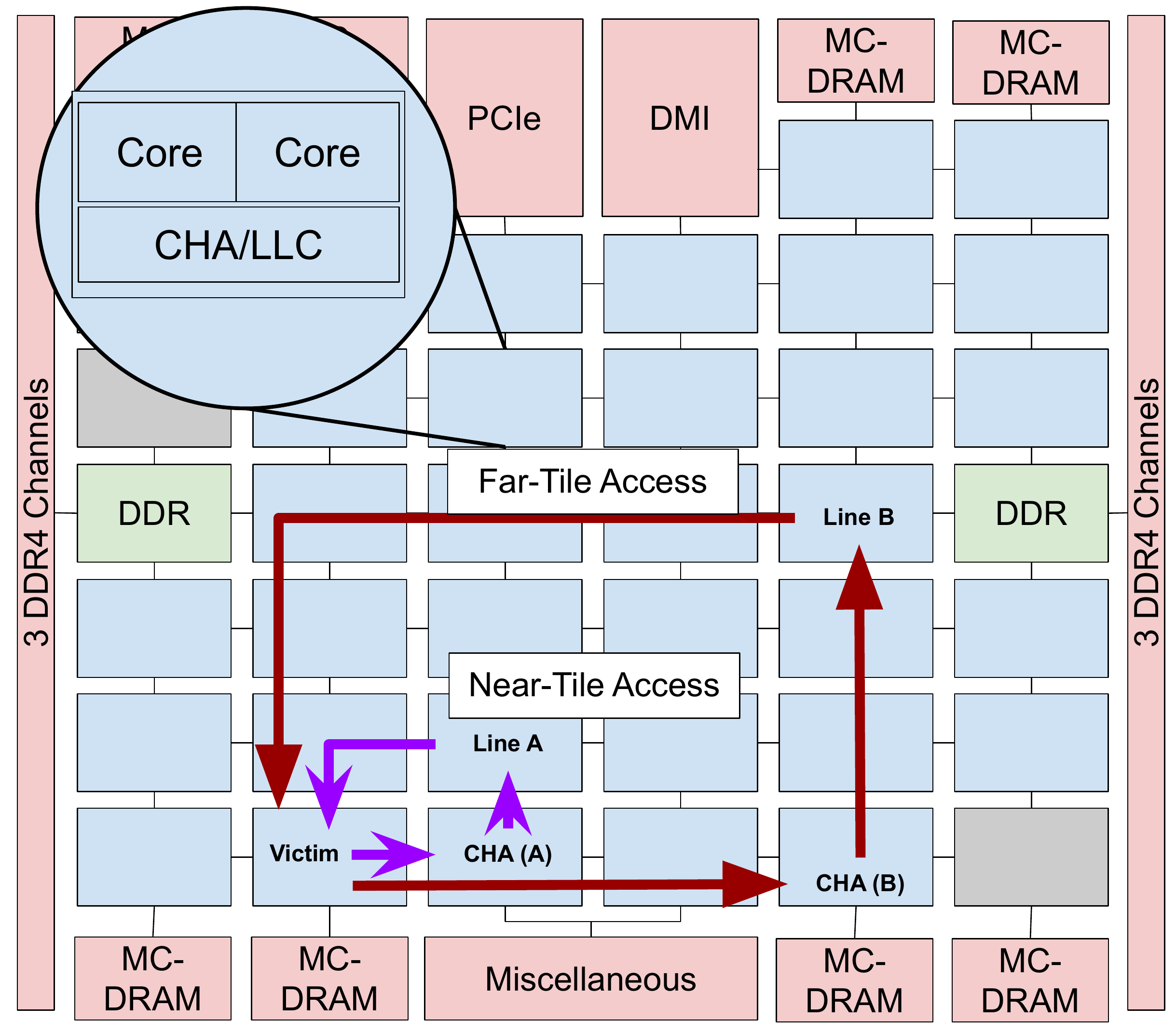}
    \caption{Knights Landing Floorplan Block Diagram~\cite{horro2019effect}. 
    Blue rectangles denote active tiles. 
    Grey rectangles denote tiles with disabled cores. One tile is zoomed to show that it contains two cores and a CHA/LLC}
    \label{fig:knl-block-diagram}
\end{figure}
\subsection{LLC Organization}
\label{sec-llc-org}
The CPU has a floorplan shown as Figure~\ref{fig:knl-block-diagram}, where its 72 physical cores. 
(or 288 physical threads with hyperthreading) are distributed across 38 tiles~\cite{horro2019effect,sodani2015knights}. 
It is known that not all tiles have active physical cores on them, and the physical CPU IDs---the IDs which are typically obtained through the Advanced Configuration and Power Interface (ACPI) and are recognized by OS---are arbitrarily assigned to a tile in an order which tends to alternate between the four quadrants.
For our attack, we do not need to explicitly know the tiles mapping, however this is obtainable using methods shown in previous works~\cite{horro2019effect}.

\subsubsection{MESIF Coherence Protocol}
The CPU employs a directory-based cache coherence mechanism using the MESIF protocol~\cite{goodman2004mesif} with a distributed directory system.
Each tile includes a Caching/Home Agent (CHA) in charge of a portion of the directory.  
Two CPUs sharing the same tile use the same CHA and LLC.
However, each CPU has its own private L1 cache. 
Since, ~\cpumodel{} is configured in flat memory mode, L2 acts as the private (but shared among two cores on same tile) last level cache. 

\subsubsection{Resolving a LLC Request}
Each time a core requests a cache line due to an L1 miss, a corresponding CHA (distributed tag directory) is queried based on the line address. 
If the cache line is present in the LLC bank of a tile, the CHA will instruct the tile to forward the data to the requester. 
Thus, two sources of latency contribute to the difference in LLC hit times - one due to difference in distance to the CHA location, and the other due to difference in distance to the forwarding tile. 
Even if two cache lines reside in the same forwarding tile, their LLC hit times can differ if two different CHAs handle the cache lines.

\subsubsection{Clustering Mode}

There are different clustering modes of the ~\cpumodel{} including  All-to-All, Quadrant Clustering and Sub NUMA Clustering (SNC) modes.
\begin{enumerate}
    \item \textbf{All-to-All} In this mode, the overall address space is uniformly distributed using the hash function across all the tiles. So, an LLC request originating from any tile can have the required CHA in any of the other tiles. There are also no restrictions on which memory controller can access which set of tiles. 
    \item \textbf{Quadrant/Hemisphere} In this mode, the memory request may originate from any tile. However, the tag directory \textit{must} be located in the same quadrant/hemisphere of the memory controller. There exists a single address space shared across all the memory channels. 
    \item \textbf{Sub NUMA Cluster (SNC-2/SNC-4)} This mode converts the processor to a 4 node NUMA(SNC-4) or 2 node NUMA(SNC-2) configuration. Each cluster has its address space interleaved within the quadrant (in SNC-4) or the hemisphere (in SNC-2).
    Memory request originating from the same cluster is served by the tag directory, memory controller within the same quadrant/ hemisphere.
    NUMA aware software can use this mode where the  each thread is pinned to a specific hemisphere(2)/quadrant(4). 
\end{enumerate}

We configure to an \textit{All-to-All} cluster mode where a request can traverse the entire mesh to contact the tag directory, then forward it to the proper memory controller to fetch the required cache line.

\section{Assumptions \& Threat Model}
\label{sec:attack-assumptions}

In this paper, we assume that the attacker's target is a NUCA architecture with a multi-hop mesh interconnection network. 
The attacker can identify vulnerable access patterns either through profiling, if they do not have access to the source code, or by recognizing it within the source code itself. 
Unlike other side-channel attacks based on contention within the NUCA architecture, the distance-based attack requires measuring the execution time of the victim functions that are known to access data in different cache banks in NUCA. 
The attacker can time the sensitive operations in the victim program by interacting with the victim, such as exchanging messages through the network or inter-process communication (IPC), detecting changes in shared variables, or exploiting other side-channels. 
In the case of using shared variable, attacker needs to have read permission on the shared variable through some libraries (like shared AES library) similar to previous attack~\cite{yarom2014flush+,gruss2016flush+,reload+refresh,coherence-protocol,lru-sidechannel}.

Restrictive defense mechanisms like not allowing hyperthreading of processes of different security domains can be assumed~\cite{cross-vm,best-practice-azure}. 
However, attacker needs to be able to launch threads to different physical cores other than the victim's core. 
For the covert channel, we assume that the attacker (spy) and the victim process are strictly cooperating, so there is no need for a separate timer thread, rather the attacker is allowed to read timestamps directly by the victim process.
For the side-channel, we assume that attacker do not have the permission to insert RDTSC or any other timing mechanism to directly monitor the victim process. 

The attackers may or may not have access to an accurate timing function like RDTSC, but they can use alternatives such as counting threads. It is also assumed that the attacker does not possess root privileges.
The hardware and the OS are assumed to be correct and trusted by the victim program. 

\section{Implementation}
\label{sec:attack}

In this section, we describe the setup and steps for exploiting a NUCA distance-based side-channel.
First, we show the attack on a simulated machine using the Gem5 simulator~\cite{binkert2011gem5}. We did this to demonstrate the attack in a strictly controlled environment where it is easier to show the main components of an attack. 
Then, we demonstrate the attack example using the AES decryption function in a real machine.
The attack is demonstrated on an \cpumodel{} CPU.

\subsection{A NUCA Distance-based Side-channel in a Simulated Machine}
\label{sec:simulated}
To demonstrate the attack, we first explain it using a simulated machine in Gem5~\cite{binkert2011gem5}. 
To perform the attack, we need to identify two addresses that have a significant difference in LLC hit times. 
To accomplish this objective, we configure an \texttt{8x8} tiled architecture with 64 tiles. 
Each tile contains a core, a private L1I and L1D cache, and a shared LLC (L2) bank. 
The size of a cache line is 64B. 
Each L1D is a 2-way associative 4kB cache. 
Each LLC bank is an 8-way associative 2MB cache. 
64 tiles are connected using an $8\times8$ mesh network. 
With this configuration, both the L1D cache and LLC use the bit[10:6] of the physical address for indexing. 
LLC bank is selected using bit[12:6] of the physical address.

An example attack code is shown in Figure~\ref{fig:code}. 
Before performing the attack, we allocate a large array (line 1 in Figure~\ref{fig:code}) and start accessing all cache lines belonging to the array (line 12-13). 
At the end of those accesses, we can infer that the initial entries of L1D cache will be evicted but they will still reside in LLC. 
This step is similar to {\em priming} the cache as used in prior cache side-channel attacks~\cite{osvik2006cache}. 
Next, we need a victim function that accesses two addresses (depending on the secret) such that the addresses reside in two different LLC banks. 

For determining which addresses would be useful for the victim function, we analyze the LLC hit latency for all the entries of the array indexed from 0 to \texttt{511*64}. 
We found that the entry at the index \texttt{117*64} and \texttt{118*64} are mapped to the virtual address \texttt{0x4c7fc0} and \texttt{0x4c8000}, accordingly. 
These virtual addresses are mapped to the physical addresses \texttt{0xc6fc0} and \texttt{0xc7000}, respectively. 
Using the LLC bank selection bits, we can verify that entry at index \texttt{117*64} (virtual address \texttt{0x4c7fc0}, physical address \texttt{0xc6fc0}) is mapped to bank 63. 
On the other hand, the entry at index 118*64 is mapped to bank 0. 
With this mapping, we make sure that the victim function accesses either index \texttt{117*64} or \texttt{118*64} depending on the secret (Lines 5-6). 
As a result, when an attacker times the victim function (Lines 16-18), he/she can infer the secret one bit at a time by comparing the access latency with the access latency of bank 63 (or 0).

\begin{figure*}[h]
    \centering
    \includegraphics[width=\columnwidth]{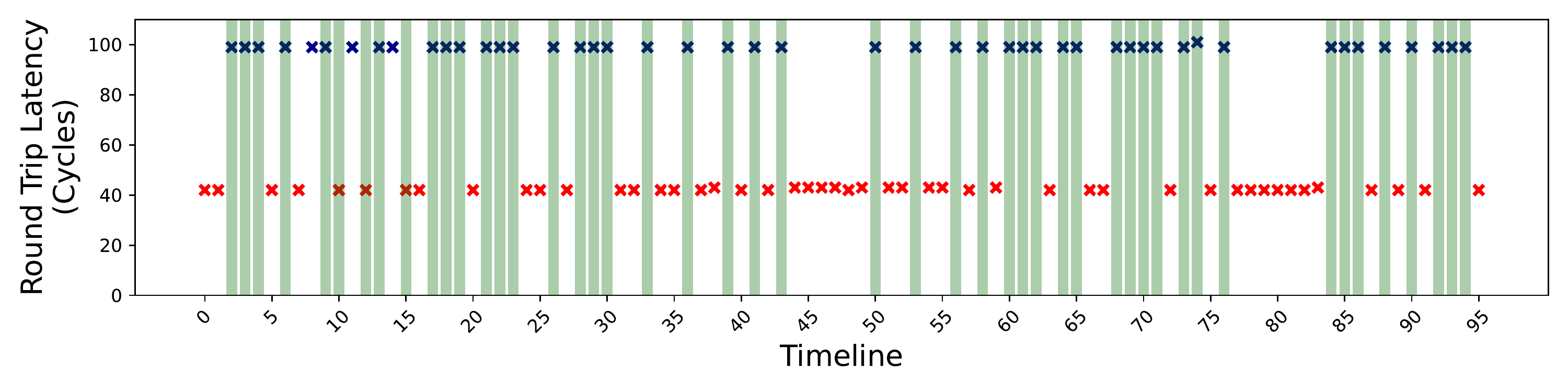}
    \caption{Round trip latency obtained from POC attack on a simulated machine. Green Region represents cases when the secret bit is 0, white region represents cases when the secret bit is 1. Blue Cross ({\color{blue}x}) represent data predicted to be bit 0 and red cross ({\color{red}x}) represent data predicted to be bit 1. We have >=95\% accuracy with other cores running Rodinia benchmark~\cite{che2009rodinia} applications}.
    \label{fig:poc}
    \vspace{-2pt}
\end{figure*}

In this run, we have the attacking code running one of the 64 cores and the rest of the cores are running instances of Rodinia v3~\cite{che2009rodinia} benchmark applications.
This was done to make sure the attack is noise resilient. 
From  Figure~\ref{fig:poc}, we can clearly see that in the case of the green area (i.e. secret bit 0), the round trip latency of the load is mostly very high (95+ cycles). On the other hand when the secret bit is 1, we have round trip latency that is very low (around ~40 cycles). In this way, by observing this round trip latency, the attacker can infer the secret successfully with high accuracy (>95\%).
%

%

\begin{figure}[t]
\begin{lstlisting}[style=CStyle,language=C]
uint8_t arr[512 * 64], secret = 123;

void victim(unsigned int mask) {
    uint8_t s = secret & mask;
    if (s == 0) s ^= arr[117 * 64]; // LLC bank 63
    else        s ^= arr[118 * 64]; // LLC bank 0
}
int main() {
    unsigned long t1, t2, junk, mast = 1;
    for (int i = 0; i < 8; i++, mask <<= 1) {
        // Bring arr[117] and arr[118] to LLC
        for (int j = 0; j < 512; j++)
            junk ^= arr[j * 64];
        _mm_mfence();
        // Time the victim function
        t1 = __rdtscp(&junk);
        victim(mask);
        t2 = __rdtscp(&junk) - t1;
        // If the bit is 0, the latency > 100
        printf("BIT[%d]: %d\n", i, t2 > 100? 0:1);
    }
}
\end{lstlisting}
\vspace{-0.2cm}
\caption{An example attack code to leak a secret. The secret data is accessed inside the victim function.}
\label{fig:code}
\end{figure}



\subsection{Attack on \cpumodel{}}
\label{sec:realattack}
In this subsection, we describe the proof-of-concept (POC) implementation of our attack on the ~\cpumodel{} machine. To realize this attack, we need to identify Far-- and Near-- tile accesses and separate them. Moreover, we need to ensure LLC hits while having L1D misses.  
\subsubsection{Identifying Far- and Near-Tile Accesses}


To perform the attack, the attacker needs to identify addresses that are mapped to the CHA on a far tile or a near tile.
To do that we use a strategy, called {\em Execute  and Profile}.
This strategy has two steps. 
First, it requires an attacker process to execute on the same tile (not necessarily the same physical core or thread) with the victim program.
The attacker process then allocates a certain amount of virtual memory and accesses it.
The attacker process uses a helper thread that accesses the addresses first to bring them to the LLC. 
The CPU tile of this LLC acts as the forwarding tile. 
When another thread of the attacker process accesses those addresses, LLC hits occur. 
Depending on the distance of the CHA associated with an address, different addresses have different LLC hit times. 
The attacker process profiles the hit times of different virtual addresses.
Based on the LLC hit times, the attacker can identify two sets of virtual addresses - one that is mapped to the far tile's CHA and one that is mapped to the near tile's CHA. 
Let us denote these two sets as $\text{VA}_\text{far}$ and $\text{VA}_\text{near}$ respectively.

\begin{figure*}[htb!]
    \centering
    \includegraphics[width=1.1\columnwidth]{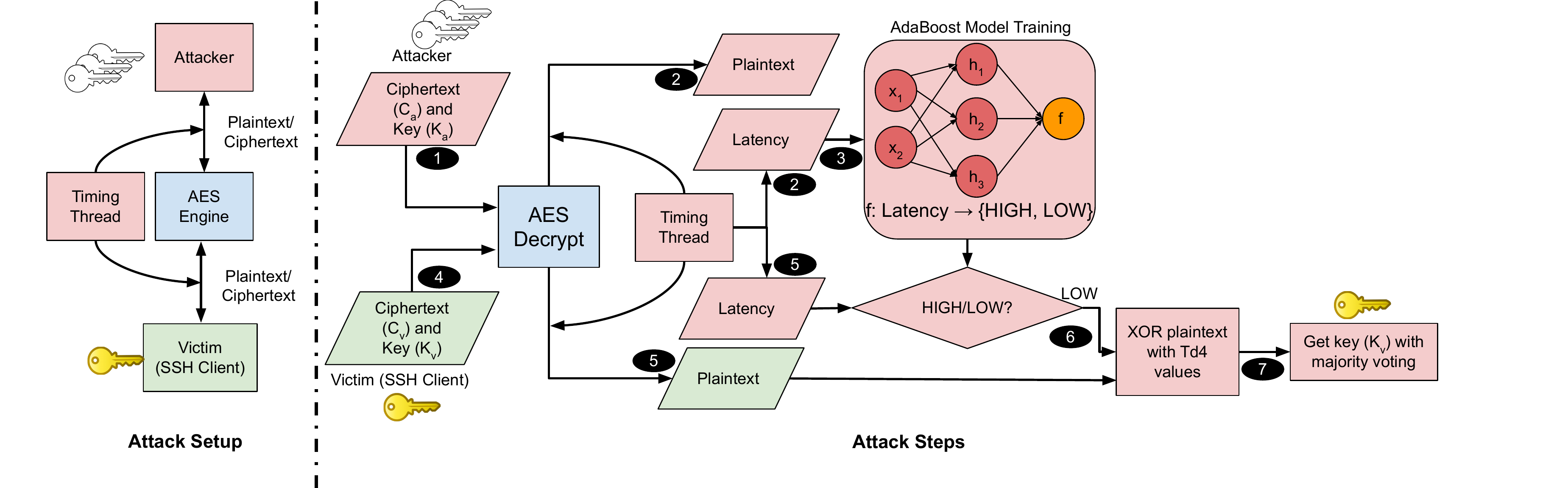}
    \caption{A step-by-step illustration of how an attacker can launch a NUCA distance-based side-channel attack on an AES code. The left side shows the setup assumed in the attack scenario.}
    \vspace{-0.5cm}
    \label{fig:attack-flowchart}
\end{figure*}
\subsubsection{Ensuring L1D Miss and LLC Hit}

The attack also requires the accessed data to be a miss in L1D but a remote hit in LLC. If the access is an L1D hit or an LLC Miss, the latency will not depend on the distance in the NoC. Also, if the access is a hit in the local LLC, no communication with the CHA will happen.
We guarantee the above scenario using two strategies:
%
%
%
\textit{First}, 
to ensure a remote LLC hit,
the attacker can run on a separate core before the victim to load the target cachelines into a remote LLC. 
Once the cachelines are loaded, other cores will continue to forward from the same LLC slice upon L1D misses.
\textit{Second}, the attacker can either prime the L1D cache of the victim by running a thread on the victim core, or invalidating the L1D cache of the victim core using instructions like {\tt PREFETCHW}.
The latter is easier since the attacker is already running a thread on another core to keep the presence in LLC.

\subsubsection{An Attack Example with AES}

\begin{figure}[t]
\begin{lstlisting}[style=CStyle,language=C,basicstyle=\scriptsize\ttfamily\bfseries]
static const u32 Td0[256] = ...;
static const u32 Td1[256] = ...;
static const u32 Td2[256] = ...;
static const u32 Td3[256] = ...;
static const u8 Td4[256] = {
  0x52U, 0x09U, 0x6aU, 0xd5U, 0x30U, 0x36U, 0xa5U, 0x38U,
  0xbfU, 0x40U, 0xa3U, 0x9eU, 0x81U, 0xf3U, 0xd7U, 0xfbU,
  ...
};

void AES_decrypt(u32 *in, u32 *out, u32 *rd_key) {
    u32 s0, s1, s2, s3, t0, t1, t2, t3;
    s0 = in[0] ^ rk[0];
    s1 = in[1] ^ rk[1];
    s2 = in[2] ^ rk[2];
    s3 = in[3] ^ rk[3];
    ...
    /* The last round */
    s0 =
        ((u32)Td4[(t0 >> 24)       ] << 24) ^
        ((u32)Td4[(t3 >> 16) & 0xff] << 16) ^
        ((u32)Td4[(t2 >>  8) & 0xff] <<  8) ^
        ((u32)Td4[(t1      ) & 0xff])       ^
        rk[0];
    PUTU32(out     , s0);
    s1 =
        ((u32)Td4[(t1 >> 24)       ] << 24) ^
        ((u32)Td4[(t0 >> 16) & 0xff] << 16) ^
        ((u32)Td4[(t3 >>  8) & 0xff] <<  8) ^
        ((u32)Td4[(t2      ) & 0xff])       ^
        rk[1];
    PUTU32(out +  4, s1);
    ...
}
\end{lstlisting}
\caption{The vulnerable, fully unrolled (i.e., non-iterative) code for AES decryption in {\tt aes\_core.c} of OpenSSL~\opensslver{}. The source code is simplified for brevity, and only shows the initial values of {\tt Td4} and the last round of {\tt AES\_decrypt}.}
\vspace{-0.2cm}
\label{fig:aes_code}
\end{figure}

AES~\cite{bonneau06aes} implementation has been used in many scenarios to authenticate users and sessions during SSH communication~\cite{ssh-protocol}. 
Here we assume the victim to be an SSH client using the AES engine of an SSH server. 
The attacker is a malicious process (possibly running on the same SSH server). 
It uses the same AES engine for encryption and decryption operations.
A \texttt{Timer} thread, orchestrated by the attacker, can time the encryption and decryption operation.
This is illustrated in Figure~\ref{fig:attack-flowchart}.

The traditional AES implementation uses a number of transformation tables, known as T tables, to represent the computation and permutation of individual bytes during multiple rounds (9 rounds for AES-128, 11 rounds for AES-192, or 13 rounds for AES-256).
These T tables have been the targets of exploitation on multiple side-channel attacks to leak the AES secret keys~\cite{Bernstein2005CachetimingAO, bonneau06aes, gorka14aesattack, gulmezoglu15aesattack}.
We show a simplified version of the AES decryption code in Figure~\ref{fig:aes_code}.
Since AES is a block cipher, in each invocation,
the {\tt AES\_decrypt} function will take a block of 128 bits as the input and decrypt it using the round key.
{\tt AES\_decrypt} and {\tt AES\_encrypt} have very similar structures, except that they use two different sets of T tables, {\tt Td0}--{\tt Td4} and {\tt Te0}--{\tt Te4}, respectively, and that {\tt AES\_decrypt} has an extra round that uses only {\tt Td4} at the end of decryption.
Generally, the last round of {\tt AES\_decrypt} has been targeted by cache side-channel attacks because, in the last round, four elements of {\tt Td4} and one element (four bytes) of the private key are XORed to produce the four bytes 
of the plaintext. For example, Line 26 - 31 does the following:

\begin{align*}
& \textsc{s0} = \textsc{Td4[A]} \wedge \textsc{Td4[B]} \wedge  \textsc{Td4[C]} \wedge \textsc{Td4[D]} \wedge \textsc{rk[1]} \\
& \textsc{PUTU32(out,s0)}    
\end{align*}

where A, B, C, and D are indices of Td4.
If the attacker knows the plaintext and Td4 values used,
then the four bytes of the private key (i.e., \texttt{rk[1]}) can be derived by XORing them. 
The key challenge is to determine which Td4 values are used here.

The NUCA distance-based side-channel attack on AES is different from the FLUSH+RELOAD and similar attacks since it cannot infer exactly which cache line is accessed 
and brought into the cache by examining the cache contents.
Instead, the attacker can only time the victim function, {\tt AES\_decrypt}, and use the latency to extract the bits inside the key.
Specifically, this attack faces two major challenges:
(1) {\bf Overlapping of multiple cache loads}: An out-of-order CPU can issue multiple load instructions into the pipeline, and send multiple requests to the Load Store Queue (LSQ). 
Although the Total Store Ordering (TSO) model of most Intel CPUs forbids reordering of the load requests, requests can still be sent while the prior requests await responses.
As a result, the latency of multiple load instructions without mutual dependency can overlap, and thus, the highest latency of individual loads will dominate the overall latency. 
(2) {\bf Timing difficulty with multiple decryption rounds}: From the attacker's point of view, it is difficult to time the last round of decryption where only elements of {\tt Td4} and the lower 4 bytes of the key are accessed. 
This is because timing the entire {\tt AES\_decrypt} function will include the time of earlier rounds of decryption making it impossible to determine how much time it takes only to access {\tt Td4} entries.

\begin{figure}[t]
    \centering
    \includegraphics[width=\linewidth]{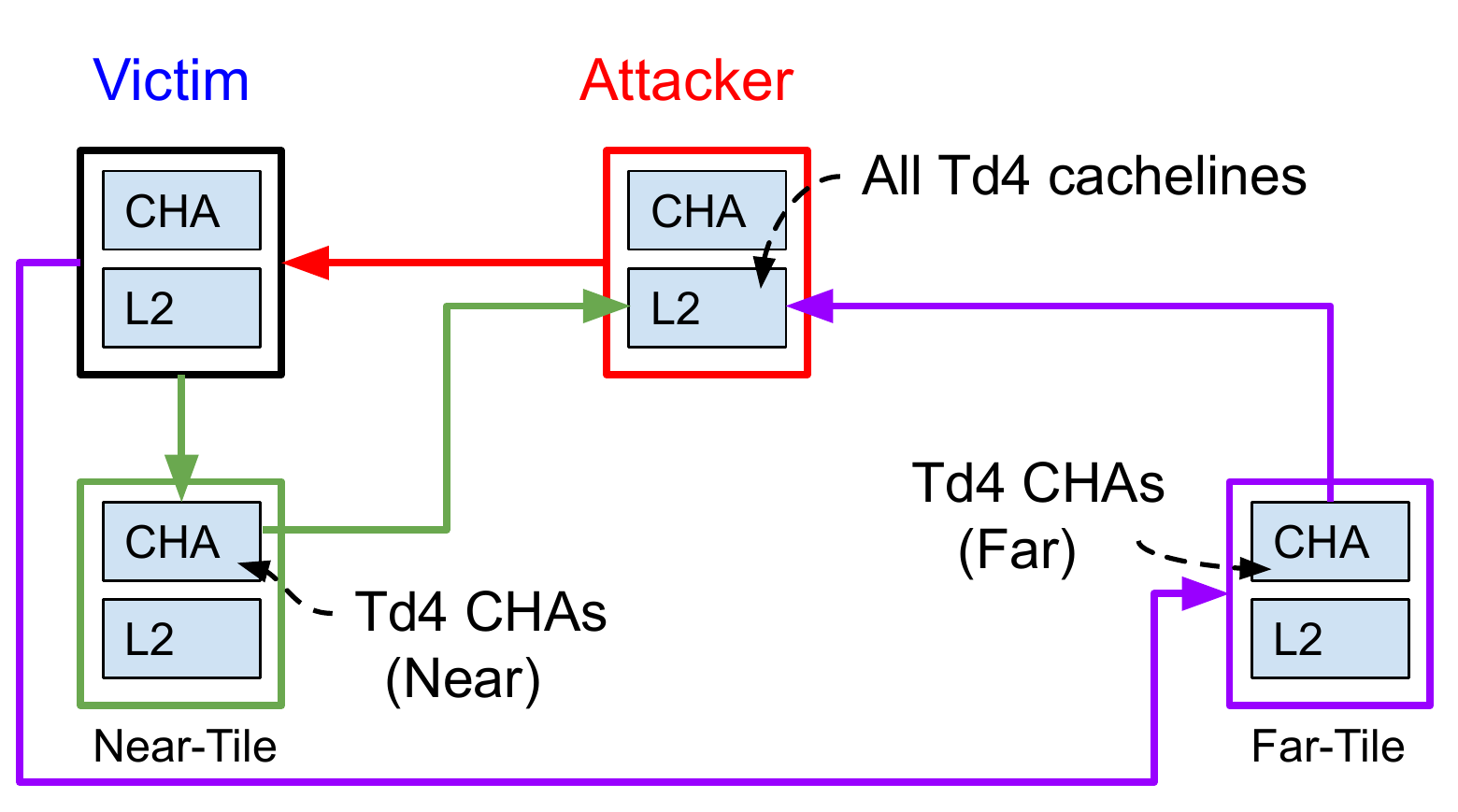}
    \caption{How the distance differences between access near-tile and far-tile CHAs are exploited by the AES attack. The attacker preloads all 4 Td4 cachelines in one core, so the LLC hit latency is determined by the distance to the CHAs.}
    \label{fig:td4}
    \vspace{-0.5cm}
\end{figure}

To overcome the challenges, we formulate the attack as follows:
The Attacker runs two threads --  thread 1 runs a loop on the same core as the AES program to bring {\tt Td0}--{\tt Td3} into L1D, while thread 2 runs on another tile to keep the whole {\tt Td4} inside LLC.
Bringing the whole {\tt Td0}--{\tt Td3} tables into the L1D is possible because the total size of {\tt Td0}--{\tt Td3} is 4KB and \cpumodel{} has 32KB per core of L1D cache. 
Any future access to {\tt Td0}--{\tt Td3} entries does not cause any network traffic in the NoC and hence, {\tt Td4} access times can be measured without any noise. We should note that {\tt Td4} table contains 4 cache lines. Although all of them will be loaded in the LLC of thread 2's tile, the CHAs of those cache lines will likely be spread apart - some of the CHA tiles will be near the victim's tile while others might be far apart as illustrated in Figure~\ref{fig:td4}. Thus, when the victim accesses Td4 table from the LLC of the attacker, some cache lines will incur lower latency while others will incur higher latency.

\texttt{Timer} thread will repeatedly check for any change in {\tt out} and {\tt out+4} (Figure~\ref{fig:aes_code}). 
We cannot rely on the timing of the whole function because there might be different LLC accesses within the AES\_decrypt call which will add to the noise.
As AES is a block cipher that uses different rounds of computation on the same table, timing the whole function complicates the secret extraction. However, by isolating the timing thread to only monitor the necessary region of interest within the function, we increase the accuracy of this attack as reflected in the results in~\S\ref{sec:experiments}. This technique can later be used to improve other cache side-channel attacks too.  
This is done through the {\tt out} buffer provided as a parameter to the {\tt AES\_decrypt} function to collect the decrypted text. 
This {\tt out} buffer is observable to the attacker.
This helps the attacker to keep track of when {\tt out} has been modified. 
The attacker can then start a timer (or start a counting loop). 

The attacker stops the timer (or terminates the loop) when {\tt out+4} is modified. 
The time between observing these two modifications to {\tt out} entries will be  directly proportional to the time for executing the statements from Line 26 to 31 in Figure~\ref{fig:aes_code}. 
In other words, this is the time for four accesses to the {\tt Td4} table (say, $T_{26-31}$).  
This method of measuring the {\tt Td4} accesses derives from the SharedArrayBuffer method proposed in~\cite{schwarz2017malware}. 
We can use PREFETCHW timer to monitor memory addresses without any write-access as shown in~\cite{adversarial-prefetch}. Results regarding using PREFETCHW timer implementation is given in~\S\ref{sec:prefetchw-timer}.
Due to the overlapping of memory loads between lines 26 and 31, 
the time $T_{26-31}$ will be lower if the CHAs used in {\tt Td4} accesses are all in near tiles or in L1D (let us call this scenario {\em LOW}) and higher if one or more accesses are to the CHAs in a far tile (let us call it {\em HIGH}). The attacker can use some simple threshold to determine whether $T_{26-31}$ can be classified as {\em LOW} or {\em HIGH}.

Figure~\ref{fig:attack-flowchart} shows the flow of our end-to-end attack.
In Step 1, an attacker randomly generates $N$ ciphertext and key pairs ($C_a$ and $K_a$) in the \texttt{Attacker} thread ($N=10M$ in our experiments). 
The attacker then uses AES Decryption to get plaintext for each round in Step 2. 
\texttt{Timer} thread measures the latency $T_{26-31}$ during each decryption. The attacker classifies the latency with labels {\em LOW} or {\em HIGH}. 
The latency numbers and their labels are used to train an AdaBoost model.
The model in essence is a more robust version of the threshold-based classification method that the attacker has already used.  


The victim SSH Client is now allowed to use the AES engine to decrypt ciphertext $C_v$ with its own secret key $K_v$ (Step 4). 
Timer thread monitors these decryptions and measures the latency in Step 5. 
In Step 6 this latency is predicted to be either {\em LOW} or {\em HIGH} using the model trained in Step 3. 
If the latency is classified as {\em LOW}, then the attacker can XOR the plaintext with Td4 values associated with the LOW label to determine a set of possible values for $K_v$ (more specifically, bytes 8-11 of $K_v$). 
The attacker repeats Step 4 - 7 multiple times and extracts possible values of $K_v$ each time. 
The attacker uses a majority voting among the possible $K_v$ values after $T$ trials. 
Figure~\ref{fig:key-extract} shows the accuracy for extracting keys after different number of trails. 
Our experiments indicate that after $T=4000$ trials, the attacker can extract 4 bytes of $K_v$ with 100\% accuracy.
\footnote{Our POC code is here - \url{https://anonymous.4open.science/r/NUCA-side-channel-6C9D}}

\subsection{Other Vulnerable Algorithms}
We can infer that our attack depends on two things - a) Data structure (buffer) that spans across multiple cachelines. and b) Secret dependent access to those cachelines. 
Due to CPU interleaving, different entries of these large data structures are allocated to different LLC slices and that's why we have secret dependent different latencies which can be used as a source to leak those secrets. 
Based on these two properties, we have explored the OpenSSL implementation of some popular cryptography algorithms from ~\cite{abood2018survey} and found many algorithms have these two properties and hence would be vulnerable to similar attack scenarios. 

 Camellia~\cite{moriai2005addition} and ARIA~\cite{aria} have a similar S-box structure 
 that expands to multiple cachelines,
 %
  Another variable key length cryptographic algorithm Blowfish~\cite{blowfish} has a similar data structure which spans multiple adjacent cachelines. 
 These algorithms can potentially be vulnerable to the same attack.
 
\begin{figure}[t]
\begin{lstlisting}[style=CStyle,language=C,basicstyle=\scriptsize\ttfamily\bfseries]
static const uint32_t S1[256] = {
    0x00636363, 0x007c7c7c, 0x00777777, 0x007b7b7b,
  ...
};

static void sl1(ARIA_u128 *o, const ARIA_u128 *x, const ARIA_u128 *y)
{
    unsigned int i;
    for (i = 0; i < ARIA_BLOCK_SIZE; i += 4) {
        o->c[i    ] = sb1[x->c[i    ] ^ y->c[i    ]];
        o->c[i + 1] = sb2[x->c[i + 1] ^ y->c[i + 1]];
        o->c[i + 2] = sb3[x->c[i + 2] ^ y->c[i + 2]];
        o->c[i + 3] = sb4[x->c[i + 3] ^ y->c[i + 3]];
    }
}
\end{lstlisting}
\vspace{-0.2cm}
\caption{The code implementation of ARIA in openssl~\opensslver{} shows similar data structure S1 that spans multiple cacheline as well as function sl1 that accesses one of those entries depending on the secret}
\label{fig:aria_code}
\end{figure}

\section{Experiment}
\label{sec:experiments}
In this section we first describe the experiment results from Gem5 simulation. 
Then we will show the experiment results for covert-channel using the timing difference between farthest and nearest CHAs.
Finally we will explain the results we have obtained using~\cpumodel{} to implement the attack on a real machine.

\subsection{Results from Gem5 simulation}
\subsubsection{Simulated Attack Latency Distribution}
\begin{figure}[h]
    \centering
    
    \includegraphics[width=0.9\linewidth]{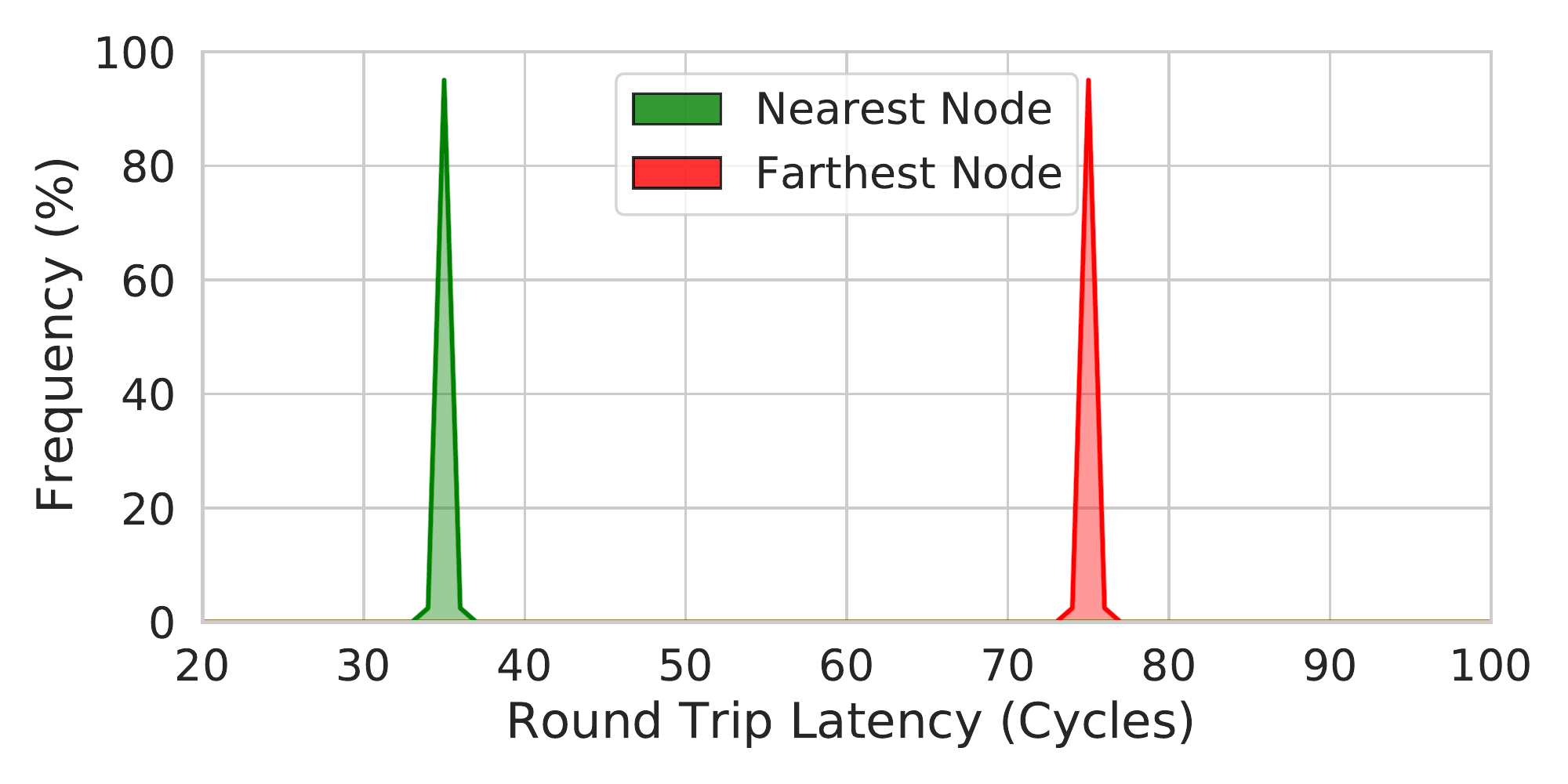}
    \caption{Simulated Attack Latency Distribution in Gem5. In this case we can clearly observe the latency difference due to the physical distance of nodes requesting the data from the source of the data}
    \label{fig:sim-attack-dist}
    \vspace{-0.2in}
\end{figure}

In Figure~\ref{fig:sim-attack-dist} we can that the latency distribution for LLC hits from the farthest and the nearest nodes. 
Timing distribution is clearly separable for LLC hit at the Near and the Far node. 
This clearly shows that the possibility of this side-channel to extract secrets.  

With a specific set of machine where we can replicate similar behavior, we can easily extract the secrets by making sure that either near or far node is hit at LLC during the cache access. 

\subsection{Covert-channel Experiments}
In this section, we show the results that we have obtained by setting the attacker and victim thread in the same process. In a covert channel setup, the attacker and victim are trying to communicate covertly. This gives us the opportunity to have synchronization between attacker and victim process. 

\subsubsection{Bandwidth \& Error Rate}
Running a similar proof of concept code with all the optimizations and steps mentioned in  Section~\ref{sec:attack} we get an accuracy of $~99.98\%$ on extracting secrets after 100,000 trials. \sk{The sentense looks broken}In these trials, the channel could reach a bandwidth of $~205$KBPS while maintaining an Error Rate of less than 0.02\%. 

\begin{figure}[h]
    \centering
    \includegraphics[width=0.9\linewidth]{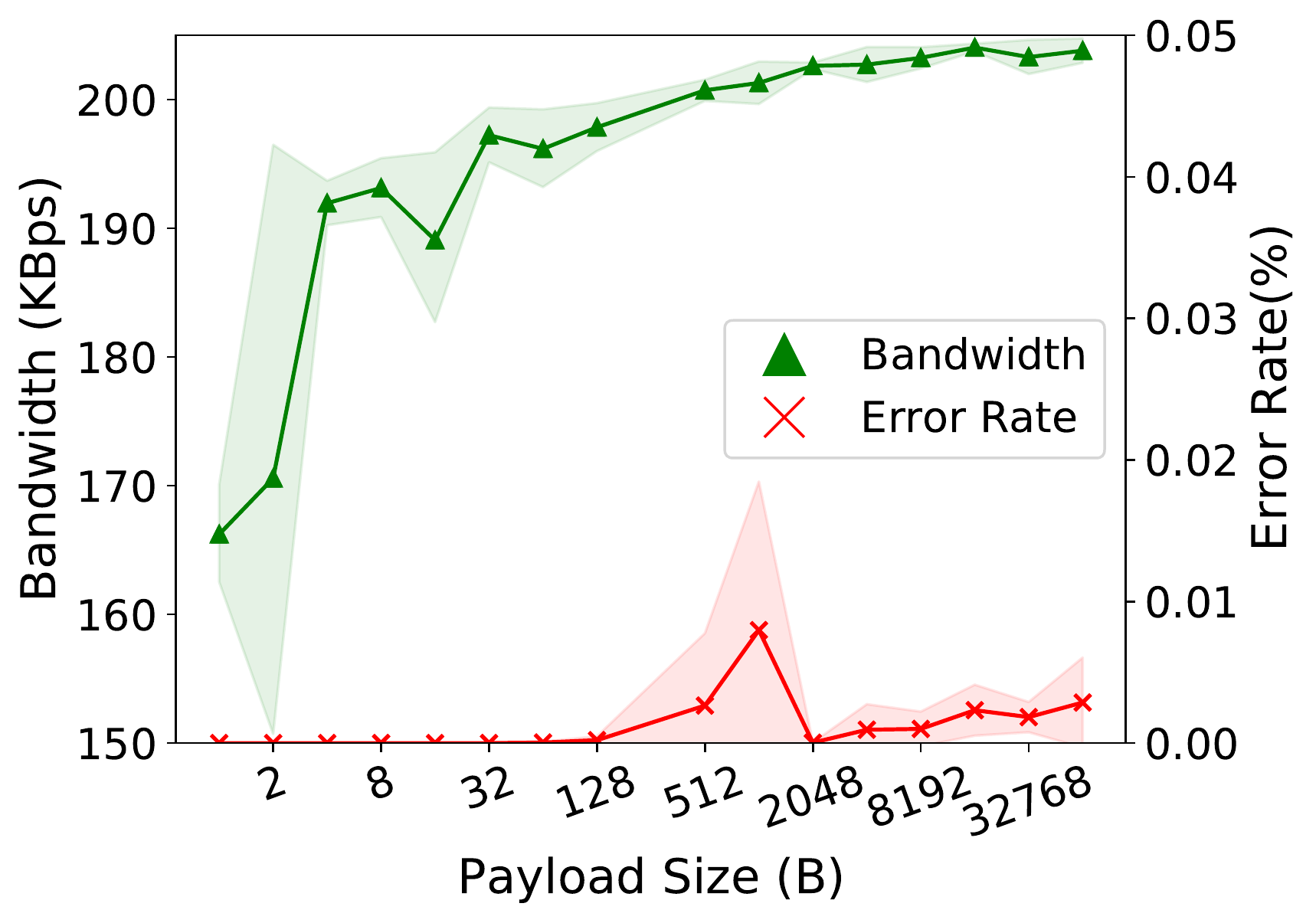}
    \caption{Bandwidth \& Error Rate of covert channel with varying payload size}
    \label{fig:bandwidth-error-rate-payload}
    \vspace{-0.1in}
\end{figure}
Figure ~\ref{fig:bandwidth-error-rate-payload} shows the Bandwidth and Error Rate of the covert channel implementation with varying size of payload. The shaded area shows 95\% confidence interval.

\subsection{Side-channel Results}
\subsubsection{PREFETCHW Timer}
\label{sec:prefetchw-timer}
\begin{figure}[h]
    \centering
    \includegraphics[width=0.9\linewidth]{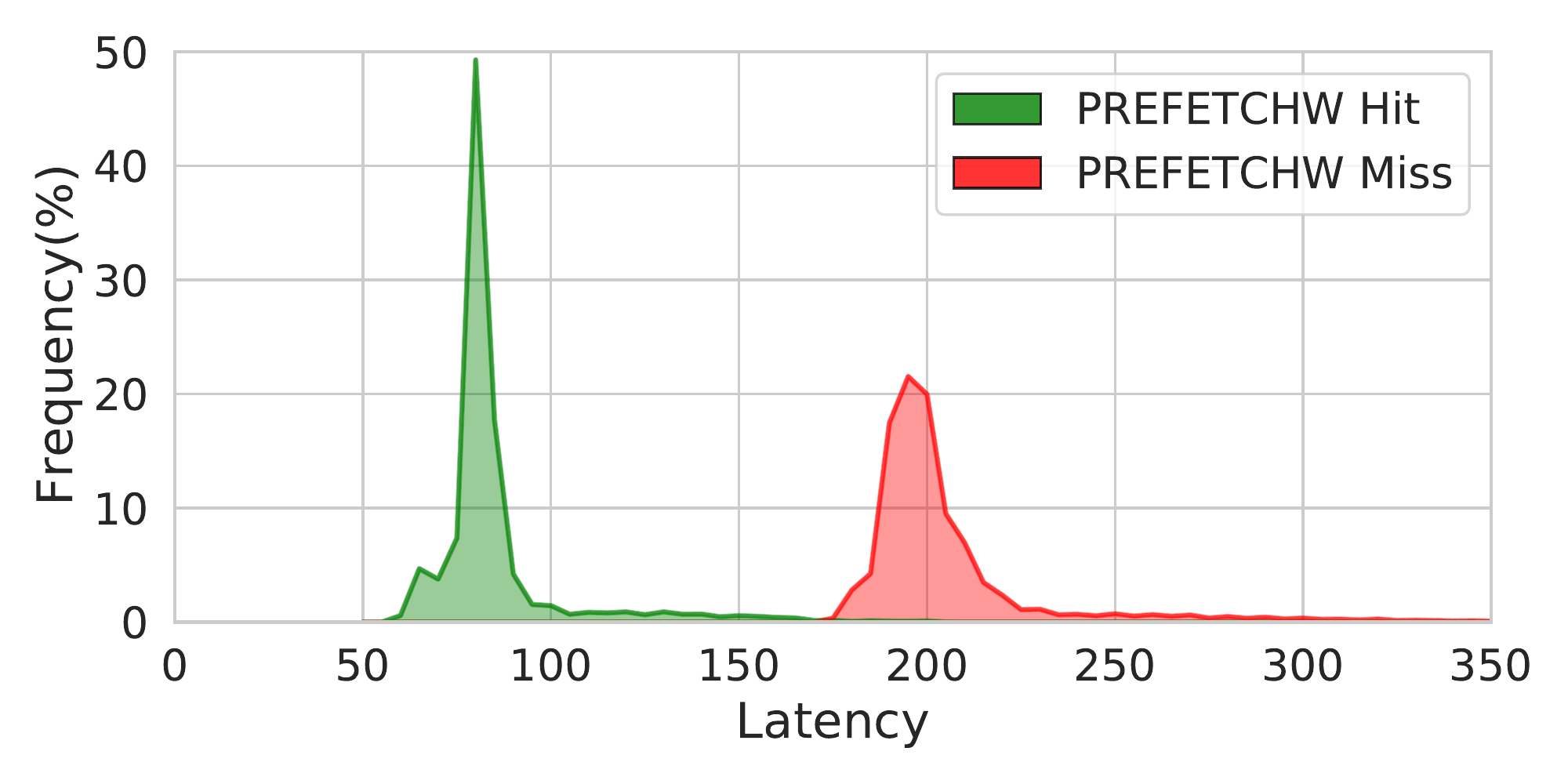}
    \caption{Hit or Miss latency distribution using PREFETCHW instructions. PREFETCHW takes much longer time to complete if a remote LLC modifies the data due to invalidations. To monitor a shared variable, we can utilize this difference.}
    \label{fig:prefetch-timer}
    \vspace{-0.1in}
\end{figure}

In this section, we discuss the possibility of using PREFETCHW for observing any change in the data following the examples of Adversarial Prefetch~\cite{adversarial-prefetch}. We used a spy thread to continuously execute PREFETCHW on a read-only memory address. When the victim thread who has write-access to that data, writes new data on that address, PREFETCHW takes much longer time >150 cycles to complete compared to <100 cycle time as observed in this experiment. The result latency distribution is shown in Figure~\ref{fig:prefetch-timer}
We can use this prefetch timer to identify whether \textit{out} shared variable was modified by the victim or not assuming that the attacker do not have write-access to the \textit{out} variable. 

\subsubsection{Accuracy of ML Classifier}
\label{sec:accuracy-ml-classifier}
\begin{figure}[h]
    \centering
    \includegraphics[width=0.85\linewidth]{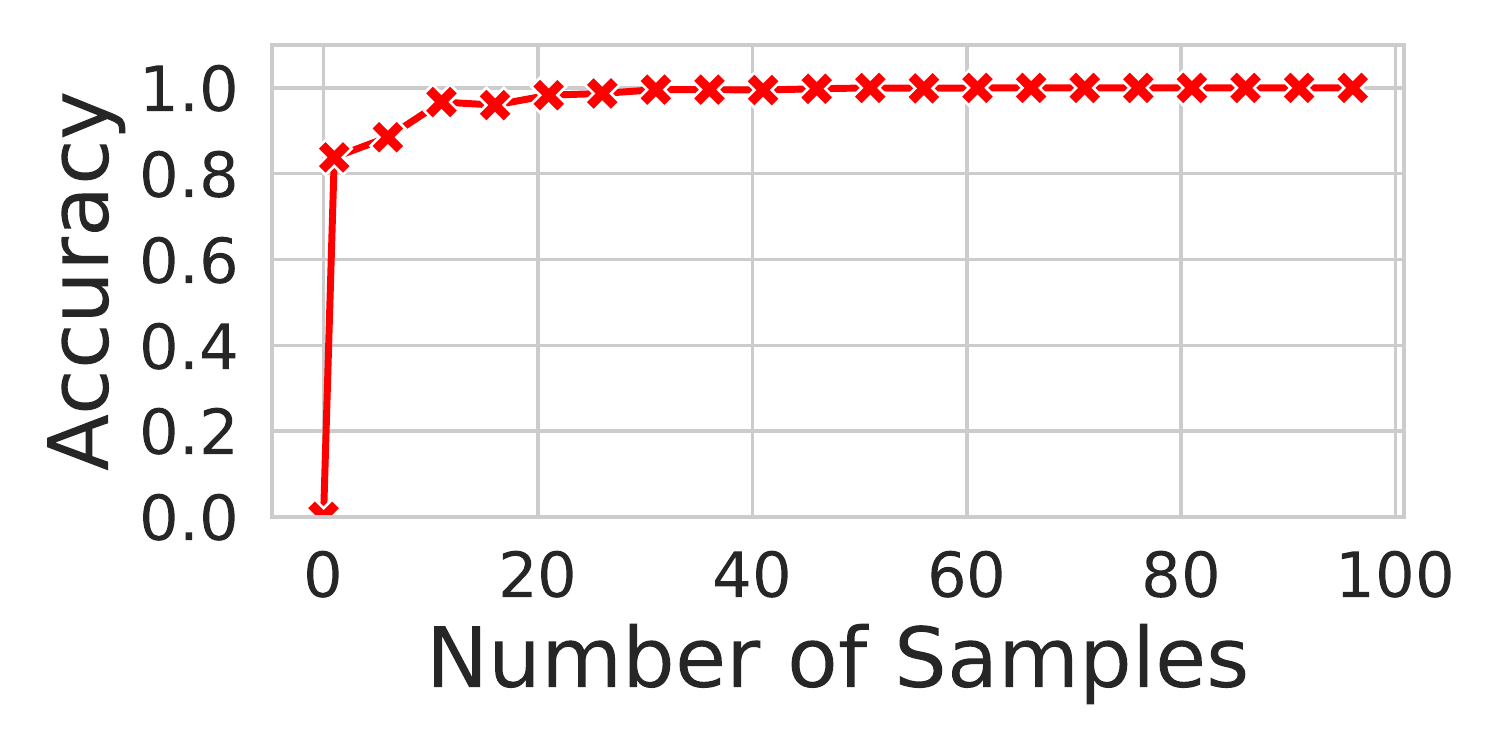}
    \caption{Accuracy of determining if one or more accesses fall to the far tile.
    We reach 100\% accuracy by taking majority voting of 40 samples or more.}
    \label{fig:ml-class-acc}
\end{figure}

In our proof-of-concept (POC) code, we take multiple samples of $T_{24-28}$ from decrypting a specific 16-byte plaintext and use majority voting (using AdaBoost Classifier~\cite{adaboost}) to determine if one or more accesses fall to a far tile. Figure~\ref{fig:ml-class-acc} shows that with AdaBoost Classifier, we can determine accesses to a far tile with 100\% accuracy by using 40 or more samples. If one or more accesses happen to a far tile, we determine the potential lower 4 bytes of the key by XOR'ing. We keep doing this using random plaintexts and eventually, extract the lower 4 bytes of the key using a simple majority for each byte.

\subsubsection{Key Extraction Accuracy}

\begin{figure}[ht]
    \centering
    \includegraphics[width=0.9\linewidth]{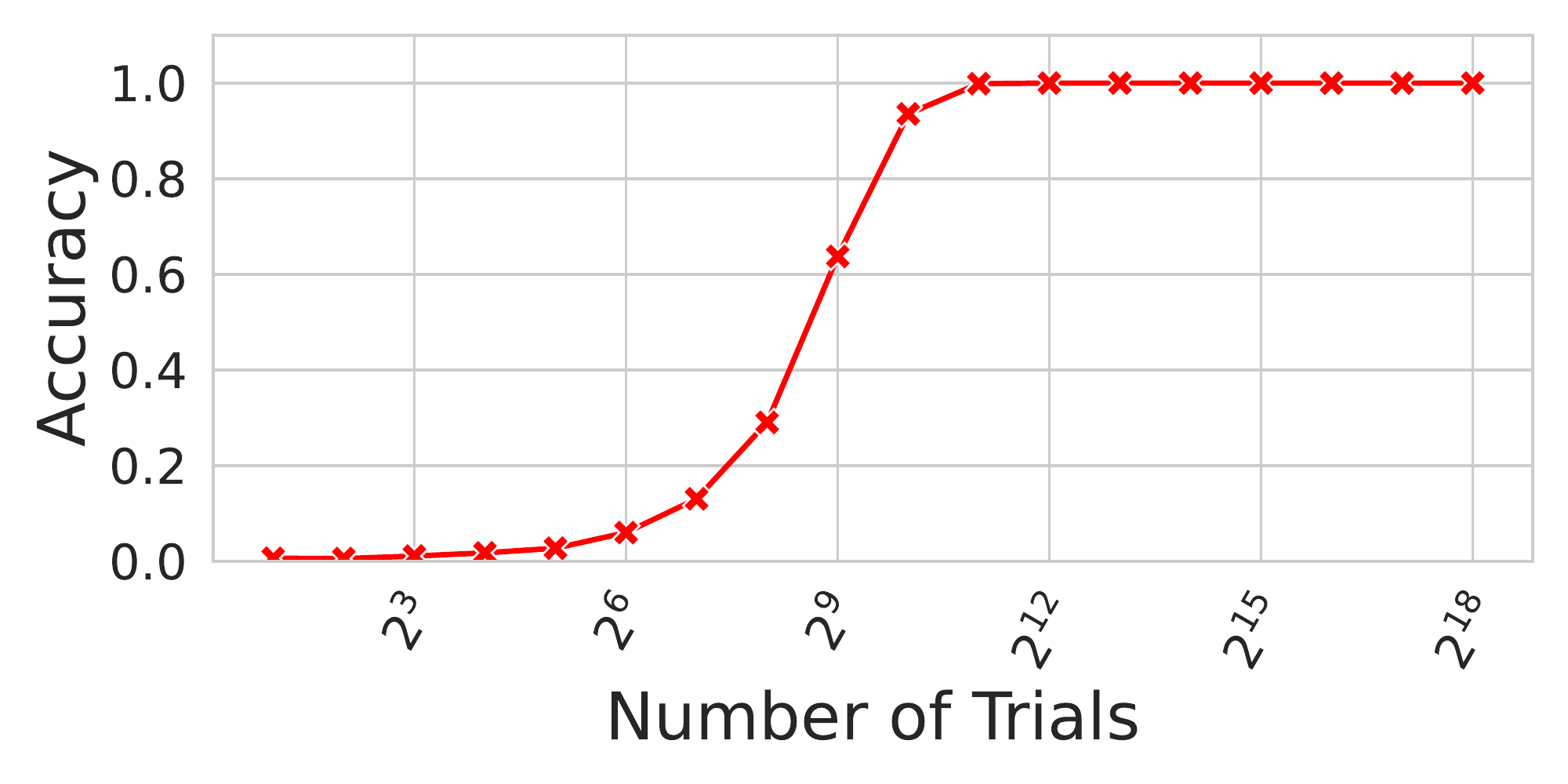}
    \caption{Key extraction accuracy with repeated decryption trials. We can extract 4 bytes of any random key with 100\% accuracy 
    by using only $\simeq 4000$ trials.}
    \vspace{-0.5cm}
    \label{fig:key-extract}
\end{figure}

Figure~\ref{fig:key-extract} shows the accuracy for extracting keys with varying number of trails. Our experiments indicate that after $T=4000$ trials, the attacker can extract 4 bytes of $K_v$ with 100\% accuracy.

\section{Possible Defense}
\label{sec:possible-defense}
In this section, we discuss about the possible defense mechanism against NUCA distance-based side-channel attacks.
\textit{First,} we need to identify the
root cause of the attack and then we can add components/techniques
to disable it. However, in the case of this attack, the source of the side-channel is because we have differences in the physical location of different CHAs representing different parts of the address space.
In the case of on-chip mesh networks, we have to take this physical distance into account while traveling hop by hop from the source node to the destination.
With larger on chip networks like~\cpumodel{} the number of hops to reach the farthest node only increases compared to the nearest nodes. 

\begin{figure}[h]
    \centering
    \includegraphics[width=0.75\linewidth]{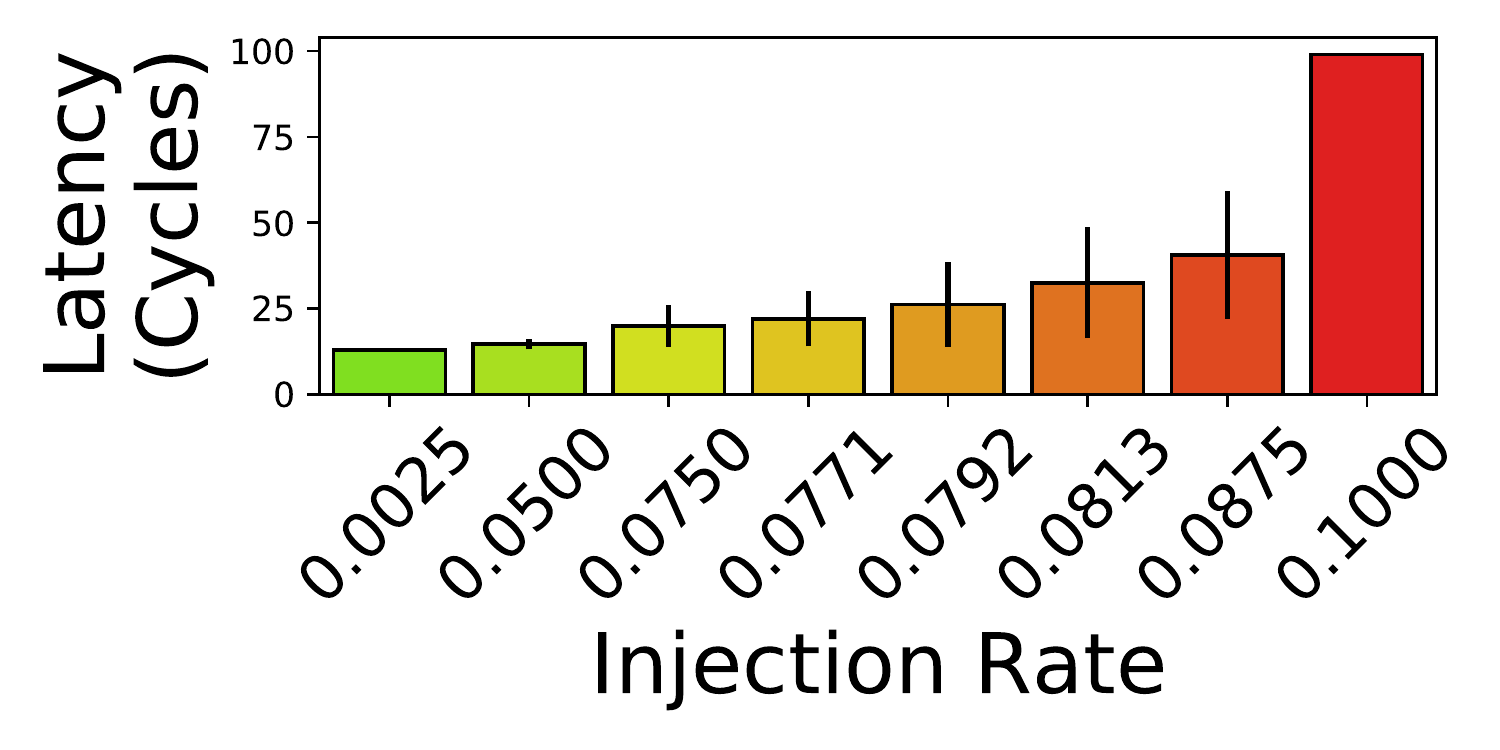}
    \caption{8x8 on chip network simulated in BookSim~\cite{booksim} with uniform random traffic. The graph shows packet network latency with varying injection rate until the network gets saturated.}
    \label{fig:network-simulation}
\end{figure}

To have an architectural defense from this type of side-channel attack, we need to make sure that all the nodes are reachable within the same amount of time. 
This can be done in many network configurations like Ring-based network where with some modification we can guarantee all the nodes can be reached within the same amount of time. 
Or, we can artificially add delays to the shorter path communication so that the attacker cannot infer the difference from timing the near and far node communications.     
In this way, all the nodes, regardless of their distance from the source, will send their response in the worst-case scenario. 
This will also harshly impact the performance. 
Our experiment results using the Booksim~\cite{booksim} simulator show that in $8\times8$ configuration, we have the packet network latency getting saturated at >100 cycles at injection rate of 0.1. 
The experiment results can be seen here in Figure~\ref{fig:network-simulation}.

As we can observe from the Figure~\ref{fig:network-simulation} at a higher injection rate, the network gets saturated. So at this latency (>100 cycles), the latency observed by the attacker would not be correlated with the distance from the attacker node, but rather be compounded by the saturated network. Following this, we can design a defense mechanism that can work against NUCA distance-based side-channel attacks. 

\section{Conclusion}
We have shown a novel NUCA distance-based side-channel attack in a simulated as well as a real machine. We demonstrated how to use this attack to break AES cryptographic algorithm in a ~\cpumodel{} machine.
We used a combination of microarchitectural techniques with machine learning to overcome major challenges like identifying overlapping access while measuring access latencies of only region of interest of a victim function. 
With Gem5 simulator~\cite{binkert2011gem5} we have shown the principle of this attack on the presence of other background application.
We have also shown how a covert channel can be created using the same principle of non uniform distance between LLC banks. 
Our covert channel can transmit information at a rate of 205KB/s with a very low error rate of 0.02\%. 
Finally, using the restricted environment, in a side-channel setting, we could extract 4 bytes of the AES key with only 4,000 decryption trials. %
We leave the future works for extracting rest of the secret key by extending this or by using some other side-channel attacks.

\bibliographystyle{acm}
\bibliography{ref}

\begin{thebibliography}{10}

\bibitem{amdryze}
Amd ryzen.
\newblock \url{https://www.amd.com/en/products/cpu/amd-epyc-7742}.

\bibitem{ampere-altra-review}
Ampere altra review.
\newblock \url{https://www.anandtech.com/show/16315/the-ampere-altra-review/3}.

\bibitem{xeonphi}
Intel xeon phi.
\newblock
  \url{https://ark.intel.com/content/www/us/en/ark/products/series/75557/intel-xeon-phi-processors.html}.

\bibitem{abood2018survey}
{\sc Abood, O.~G., and Guirguis, S.~K.}
\newblock A survey on cryptography algorithms.
\newblock {\em International Journal of Scientific and Research Publications
  8}, 7 (2018), 495--516.

\bibitem{aciiccmez2008vulnerability}
{\sc Ac{\i}i{\c{c}}mez, O., and Schindler, W.}
\newblock A vulnerability in rsa implementations due to instruction cache
  analysis and its demonstration on openssl.
\newblock In {\em Cryptographers’ Track at the RSA Conference\/} (2008),
  Springer, pp.~256--273.

\bibitem{cascadelake}
{\sc Arafa, M., Fahim, B., Kottapalli, S., Kumar, A., Looi, L.~P., Mandava, S.,
  Rudoff, A., Steiner, I.~M., Valentine, B., Vedaraman, G., et~al.}
\newblock Cascade lake: Next generation intel xeon scalable processor.
\newblock {\em IEEE Micro 39}, 2 (2019), 29--36.

\bibitem{Bernstein2005CachetimingAO}
{\sc Bernstein, D.~J.}
\newblock Cache-timing attacks on aes.

\bibitem{binkert2011gem5}
{\sc Binkert, N., Beckmann, B., Black, G., Reinhardt, S.~K., Saidi, A., Basu,
  A., Hestness, J., Hower, D.~R., Krishna, T., Sardashti, S., et~al.}
\newblock The gem5 simulator.
\newblock {\em ACM SIGARCH computer architecture news 39}, 2 (2011), 1--7.

\bibitem{bonneau06aes}
{\sc Bonneau, J., and Mironov, I.}
\newblock Cache-collision timing attacks against aes.
\newblock In {\em Proceedings of the 8th International Conference on
  Cryptographic Hardware and Embedded Systems\/} (Berlin, Heidelberg, 2006),
  CHES'06, Springer-Verlag, p.~201–215.

\bibitem{borkar1999design}
{\sc Borkar, S.}
\newblock Design challenges of technology scaling.
\newblock {\em IEEE micro 19}, 4 (1999), 23--29.

\bibitem{brasser2017software}
{\sc Brasser, F., M{\"u}ller, U., Dmitrienko, A., Kostiainen, K., Capkun, S.,
  and Sadeghi, A.-R.}
\newblock Software grand exposure:$\{$SGX$\}$ cache attacks are practical.
\newblock In {\em 11th $\{$USENIX$\}$ Workshop on Offensive Technologies
  ($\{$WOOT$\}$ 17)\/} (2017).

\bibitem{reload+refresh}
{\sc Briongos, S., Malag{\'o}n, P., Moya, J.~M., and Eisenbarth, T.}
\newblock Reload+ refresh: Abusing cache replacement policies to perform
  stealthy cache attacks.
\newblock In {\em Proceedings of the 29th USENIX Conference on Security
  Symposium\/} (2020), pp.~1967--1984.

\bibitem{che2009rodinia}
{\sc Che, S., Boyer, M., Meng, J., Tarjan, D., Sheaffer, J.~W., Lee, S.-H., and
  Skadron, K.}
\newblock Rodinia: A benchmark suite for heterogeneous computing.
\newblock In {\em 2009 IEEE international symposium on workload
  characterization (IISWC)\/} (2009), Ieee, pp.~44--54.

\bibitem{aes-original}
{\sc Daemen, J., and Rijmen, V.}
\newblock Aes proposal: Rijndael.

\bibitem{dont-mess-around}
{\sc Dai, M., Paccagnella, R., Gomez-Garcia, M., McCalpin, J., and Yan, M.}
\newblock Don't mesh around:$\{$Side-Channel$\}$ attacks and mitigations on
  mesh interconnects.
\newblock In {\em 31st USENIX Security Symposium (USENIX Security 22)\/}
  (2022), pp.~2857--2874.

\bibitem{cori}
{\sc Doerfler, D., Austin, B., Cook, B., Deslippe, J., Kandalla, K., and
  Mendygral, P.}
\newblock Evaluating the networking characteristics of the cray xc-40 intel
  knights landing-based cori supercomputer at nersc.
\newblock {\em Concurrency and Computation: Practice and Experience 30}, 1
  (2018), e4297.

\bibitem{godfrey2013server}
{\sc Godfrey, M., and Zulkernine, M.}
\newblock A server-side solution to cache-based side-channel attacks in the
  cloud.
\newblock In {\em 2013 IEEE Sixth International Conference on Cloud
  Computing\/} (2013), IEEE, pp.~163--170.

\bibitem{goodman2004mesif}
{\sc Goodman, J., and Hum, H.}
\newblock Mesif: A two-hop cache coherency protocol for point-to-point
  interconnects (2004).

\bibitem{tlb-attack}
{\sc Gras, B., Razavi, K., Bos, H., Giuffrida, C., et~al.}
\newblock Translation leak-aside buffer: Defeating cache side-channel
  protections with tlb attacks.
\newblock In {\em USENIX Security Symposium\/} (2018), vol.~216.

\bibitem{cloakhtm}
{\sc Gruss, D., Lettner, J., Schuster, F., Ohrimenko, O., Haller, I., and
  Costa, M.}
\newblock Strong and efficient cache side-channel protection using hardware
  transactional memory.
\newblock In {\em USENIX Security Symposium\/} (2017), pp.~217--233.

\bibitem{gruss2016flush+}
{\sc Gruss, D., Maurice, C., Wagner, K., and Mangard, S.}
\newblock Flush+ flush: a fast and stealthy cache attack.
\newblock In {\em International Conference on Detection of Intrusions and
  Malware, and Vulnerability Assessment\/} (2016), Springer, pp.~279--299.

\bibitem{gruss2015cache}
{\sc Gruss, D., Spreitzer, R., and Mangard, S.}
\newblock Cache template attacks: Automating attacks on inclusive last-level
  caches.
\newblock In {\em 24th $\{$USENIX$\}$ Security Symposium ($\{$USENIX$\}$
  Security 15)\/} (2015), pp.~897--912.

\bibitem{aes-isa}
{\sc Gueron, S.}
\newblock Advanced encryption standard (aes) instructions set.
\newblock {\em Intel, http://softwarecommunity. intel. com/articles/eng/3788.
  htm, accessed 25\/} (2008).

\bibitem{gullasch2011cache}
{\sc Gullasch, D., Bangerter, E., and Krenn, S.}
\newblock Cache games--bringing access-based cache attacks on aes to practice.
\newblock In {\em 2011 IEEE Symposium on Security and Privacy\/} (2011), IEEE,
  pp.~490--505.

\bibitem{gulmezoglu15aesattack}
{\sc G\"{u}lmezo\u{g}lu, B., undefinednci, M.~S., Irazoqui, G., Eisenbarth, T.,
  and Sunar, B.}
\newblock A faster and more realistic flush+reload attack on aes.
\newblock In {\em Revised Selected Papers of the 6th International Workshop on
  Constructive Side-Channel Analysis and Secure Design - Volume 9064\/}
  (Berlin, Heidelberg, 2015), COSADE 2015, Springer-Verlag, p.~111–126.

\bibitem{adversarial-prefetch}
{\sc Guo, Y., Zigerelli, A., Zhang, Y., and Yang, J.}
\newblock Adversarial prefetch: New cross-core cache side channel attacks.
\newblock In {\em 2022 IEEE Symposium on Security and Privacy (SP)\/} (2022),
  IEEE, pp.~1458--1473.

\bibitem{adaboost}
{\sc Hastie, T., Rosset, S., Zhu, J., and Zou, H.}
\newblock Multi-class adaboost.
\newblock {\em Statistics and its Interface 2}, 3 (2009), 349--360.

\bibitem{catxeon}
{\sc Herdrich, A., Verplanke, E., Autee, P., Illikkal, R., Gianos, C., Singhal,
  R., and Iyer, R.}
\newblock Cache qos: From concept to reality in the intel{\textregistered}
  xeon{\textregistered} processor e5-2600 v3 product family.
\newblock In {\em 2016 IEEE International Symposium on High Performance
  Computer Architecture (HPCA)\/} (2016), IEEE, pp.~657--668.

\bibitem{hoeneisen1972fundamental}
{\sc Hoeneisen, B., and Mead, C.~A.}
\newblock Fundamental limitations in microelectronics—i. mos technology.
\newblock {\em Solid-State Electronics 15}, 7 (1972), 819--829.

\bibitem{horro2019effect}
{\sc Horro, M., Kandemir, M.~T., Pouchet, L.-N., Rodr{\'\i}guez, G., and
  Touri{\~n}o, J.}
\newblock Effect of distributed directories in mesh interconnects.
\newblock In {\em Proceedings of the 56th Annual Design Automation Conference
  2019\/} (2019), pp.~1--6.

\bibitem{cat}
{\sc Intel, C.}
\newblock Improving real-time performance by utilizing cache allocation
  technology.
\newblock {\em Intel Corporation, April\/} (2015).

\bibitem{irazoqui2014wait}
{\sc Irazoqui, G., Inci, M.~S., Eisenbarth, T., and Sunar, B.}
\newblock Wait a minute! a fast, cross-vm attack on aes.
\newblock In {\em International Workshop on Recent Advances in Intrusion
  Detection\/} (2014), Springer, pp.~299--319.

\bibitem{gorka14aesattack}
{\sc Irazoqui, G., Inci, M.~S., Eisenbarth, T., and Sunar, B.}
\newblock Wait a minute! a fast, cross-vm attack on aes.
\newblock In {\em Research in Attacks, Intrusions and Defenses\/} (Cham, 2014),
  A.~Stavrou, H.~Bos, and G.~Portokalidis, Eds., Springer International
  Publishing, pp.~299--319.

\bibitem{jaleel2006last}
{\sc Jaleel, A., Mattina, M., and Jacob, B.}
\newblock Last level cache (llc) performance of data mining workloads on a
  cmp-a case study of parallel bioinformatics workloads.
\newblock In {\em The Twelfth International Symposium on High-Performance
  Computer Architecture, 2006.\/} (2006), IEEE, pp.~88--98.

\bibitem{booksim}
{\sc Jiang, N., Becker, D.~U., Michelogiannakis, G., Balfour, J., Towles, B.,
  Shaw, D.~E., Kim, J., and Dally, W.~J.}
\newblock A detailed and flexible cycle-accurate network-on-chip simulator.
\newblock In {\em 2013 IEEE international symposium on performance analysis of
  systems and software (ISPASS)\/} (2013), IEEE, pp.~86--96.

\bibitem{nurion}
{\sc Kang, J.-S., Myung, H., and Yuk, J.-H.}
\newblock Examination of computational performance and potential applications
  of a global numerical weather prediction model mpas using kisti supercomputer
  nurion.
\newblock {\em Journal of Marine Science and Engineering 9}, 10 (2021), 1147.

\bibitem{kayaalp2016high}
{\sc Kayaalp, M., Ponomarev, D., Abu-Ghazaleh, N., and Jaleel, A.}
\newblock A high-resolution side-channel attack on last-level cache.
\newblock In {\em 2016 53nd ACM/EDAC/IEEE Design Automation Conference (DAC)\/}
  (2016), IEEE, pp.~1--6.

\bibitem{pageplacement}
{\sc Kessler, R.~E., and Hill, M.~D.}
\newblock Page placement algorithms for large real-indexed caches.
\newblock {\em ACM Transactions on Computer Systems (TOCS) 10}, 4 (1992),
  338--359.

\bibitem{kim2002adaptive}
{\sc Kim, C., Burger, D., and Keckler, S.~W.}
\newblock An adaptive, non-uniform cache structure for wire-delay dominated
  on-chip caches.
\newblock In {\em Proceedings of the 10th international conference on
  Architectural support for programming languages and operating systems\/}
  (2002), pp.~211--222.

\bibitem{kim2012stealthmem}
{\sc Kim, T., Peinado, M., and Mainar-Ruiz, G.}
\newblock $\{$STEALTHMEM$\}$: System-level protection against cache-based side
  channel attacks in the cloud.
\newblock In {\em 21st $\{$USENIX$\}$ Security Symposium ($\{$USENIX$\}$
  Security 12)\/} (2012), pp.~189--204.

\bibitem{kiriansky2018dawg}
{\sc Kiriansky, V., Lebedev, I., Amarasinghe, S., Devadas, S., and Emer, J.}
\newblock Dawg: A defense against cache timing attacks in speculative execution
  processors.
\newblock In {\em 2018 51st Annual IEEE/ACM International Symposium on
  Microarchitecture (MICRO)\/} (2018), IEEE, pp.~974--987.

\bibitem{aria}
{\sc Kwon, D., Kim, J., Park, S., Sung, S.~H., Sohn, Y., Song, J.~H., Yeom, Y.,
  Yoon, E.-J., Lee, S., Lee, J., et~al.}
\newblock New block cipher: Aria.
\newblock In {\em Information Security and Cryptology-ICISC 2003: 6th
  International Conference, Seoul, Korea, November 27-28, 2003. Revised Papers
  6\/} (2004), Springer, pp.~432--445.

\bibitem{liu2016catalyst}
{\sc Liu, F., Ge, Q., Yarom, Y., Mckeen, F., Rozas, C., Heiser, G., and Lee,
  R.~B.}
\newblock Catalyst: Defeating last-level cache side channel attacks in cloud
  computing.
\newblock In {\em 2016 IEEE international symposium on high performance
  computer architecture (HPCA)\/} (2016), IEEE, pp.~406--418.

\bibitem{liu2015last}
{\sc Liu, F., Yarom, Y., Ge, Q., Heiser, G., and Lee, R.~B.}
\newblock Last-level cache side-channel attacks are practical.
\newblock In {\em 2015 IEEE symposium on security and privacy\/} (2015), IEEE,
  pp.~605--622.

\bibitem{trinity}
{\sc Lujan, J., Vigil, M., Kenyon, G., Sanbonmatsu, K., and Albright, B.}
\newblock Trinity supercomputer now fully operational.
\newblock Tech. rep., Los Alamos National Lab.(LANL), Los Alamos, NM (United
  States), 2017.

\bibitem{best-practice-azure}
{\sc Marshall, A., Howard, M., Bugher, G., Harden, B., Kaufman, C., Rues, M.,
  and Bertocci, V.}
\newblock Security best practices for developing windows azure applications.
\newblock {\em Microsoft Corp 42\/} (2010), 12--15.

\bibitem{moriai2005addition}
{\sc Moriai, S., Kato, A., and Kanda, M.}
\newblock Addition of camellia cipher suites to transport layer security (tls).
\newblock Tech. rep., 2005.

\bibitem{skylake}
{\sc Mujtaba, H.}
\newblock Intel skylake-x and skylake-sp mesh architecture for xcc ``extreme
  core count" cpus detailed – features higher efficiency, higher bandwidth
  and lower latency.

\bibitem{muralimanohar2007interconnect}
{\sc Muralimanohar, N., and Balasubramonian, R.}
\newblock Interconnect design considerations for large nuca caches.
\newblock {\em ACM SIGARCH Computer Architecture News 35}, 2 (2007), 369--380.

\bibitem{neve2006advances}
{\sc Neve, M., and Seifert, J.-P.}
\newblock Advances on access-driven cache attacks on aes.
\newblock In {\em International Workshop on Selected Areas in Cryptography\/}
  (2006), Springer, pp.~147--162.

\bibitem{osvik2006cache}
{\sc Osvik, D.~A., Shamir, A., and Tromer, E.}
\newblock Cache attacks and countermeasures: the case of aes.
\newblock In {\em Cryptographers’ track at the RSA conference\/} (2006),
  Springer, pp.~1--20.

\bibitem{paccagnella2021lotr}
{\sc Paccagnella, R., Luo, L., and Fletcher, C.~W.}
\newblock Lord of the ring(s): Side channel attacks on the {CPU} on-chip ring
  interconnect are practical.
\newblock In {\em 30th {USENIX} Security Symposium ({USENIX} Security 21)\/}
  (Aug. 2021), {USENIX} Association.

\bibitem{percival2005cache}
{\sc Percival, C.}
\newblock Cache missing for fun and profit, 2005.

\bibitem{caesar}
{\sc Qureshi, M.~K.}
\newblock New attacks and defense for encrypted-address cache.
\newblock In {\em Proceedings of the 46th International Symposium on Computer
  Architecture\/} (2019), pp.~360--371.

\bibitem{reinbrecht2016side}
{\sc Reinbrecht, C., Susin, A., Bossuet, L., Sigl, G., and Sep{\'u}lveda, J.}
\newblock Side channel attack on noc-based mpsocs are practical: Noc prime+
  probe attack.
\newblock In {\em 2016 29th Symposium on Integrated Circuits and Systems Design
  (SBCCI)\/} (2016), IEEE, pp.~1--6.

\bibitem{amdnuca}
{\sc Research, T.}
\newblock Amd optimizes epyc memory with numa.
\newblock Available at
  \url{https://www.amd.com/system/files/2018-03/AMD-Optimizes-EPYC-Memory-With-NUMA.pdf}
  (2021/08/12), march 2018.

\bibitem{blowfish}
{\sc Schneier, B.}
\newblock Description of a new variable-length key, 64-bit block cipher
  (blowfish).
\newblock In {\em Fast Software Encryption: Cambridge Security Workshop
  Cambridge, UK, December 9--11, 1993 Proceedings\/} (2005), Springer,
  pp.~191--204.

\bibitem{schwarz2017malware}
{\sc Schwarz, M., Weiser, S., Gruss, D., Maurice, C., and Mangard, S.}
\newblock Malware guard extension: Using sgx to conceal cache attacks.
\newblock In {\em International Conference on Detection of Intrusions and
  Malware, and Vulnerability Assessment\/} (2017), Springer, pp.~3--24.

\bibitem{sodani2015knights}
{\sc Sodani, A.}
\newblock Knights landing (knl): 2nd generation intel{\textregistered} xeon phi
  processor.
\newblock In {\em 2015 IEEE Hot Chips 27 Symposium (HCS)\/} (2015), IEEE,
  pp.~1--24.

\bibitem{stampede2}
{\sc Stanzione, D., Barth, B., Gaffney, N., Gaither, K., Hempel, C., Minyard,
  T., Mehringer, S., Wernert, E., Tufo, H., Panda, D., et~al.}
\newblock Stampede 2: The evolution of an xsede supercomputer.
\newblock In {\em Proceedings of the Practice and Experience in Advanced
  Research Computing 2017 on Sustainability, Success and Impact}. 2017,
  pp.~1--8.

\bibitem{oram}
{\sc Stefanov, E., Dijk, M.~V., Shi, E., Chan, T.-H.~H., Fletcher, C., Ren, L.,
  Yu, X., and Devadas, S.}
\newblock Path oram: An extremely simple oblivious ram protocol.
\newblock {\em J. ACM 65}, 4 (Apr. 2018).

\bibitem{skylake-sp}
{\sc Tam, S.~M., Muljono, H., Huang, M., Iyer, S., Royneogi, K., Satti, N.,
  Qureshi, R., Chen, W., Wang, T., Hsieh, H., et~al.}
\newblock Skylake-sp: A 14nm 28-core xeon{\textregistered} processor.
\newblock In {\em 2018 IEEE International Solid-State Circuits
  Conference-(ISSCC)\/} (2018), IEEE, pp.~34--36.

\bibitem{tromer2010efficient}
{\sc Tromer, E., Osvik, D.~A., and Shamir, A.}
\newblock Efficient cache attacks on aes, and countermeasures.
\newblock {\em Journal of Cryptology 23}, 1 (2010), 37--71.

\bibitem{van2015just}
{\sc Van~de Pol, J., Smart, N.~P., and Yarom, Y.}
\newblock Just a little bit more.
\newblock In {\em Cryptographers’ Track at the RSA Conference\/} (2015),
  Springer, pp.~3--21.

\bibitem{wang2007new}
{\sc Wang, Z., and Lee, R.~B.}
\newblock New cache designs for thwarting software cache-based side channel
  attacks.
\newblock In {\em Proceedings of the 34th annual international symposium on
  Computer architecture\/} (2007), pp.~494--505.

\bibitem{wassel2013surfnoc}
{\sc Wassel, H.~M., Gao, Y., Oberg, J.~K., Huffmire, T., Kastner, R., Chong,
  F.~T., and Sherwood, T.}
\newblock Surfnoc: A low latency and provably non-interfering approach to
  secure networks-on-chip.
\newblock {\em ACM SIGARCH Computer Architecture News 41}, 3 (2013), 583--594.

\bibitem{lru-sidechannel}
{\sc Xiong, W., and Szefer, J.}
\newblock Leaking information through cache lru states.
\newblock In {\em 2020 IEEE International Symposium on High Performance
  Computer Architecture (HPCA)\/} (2020), IEEE, pp.~139--152.

\bibitem{coherence-protocol}
{\sc Yao, F., Doroslovacki, M., and Venkataramani, G.}
\newblock Are coherence protocol states vulnerable to information leakage?
\newblock In {\em 2018 IEEE International Symposium on High Performance
  Computer Architecture (HPCA)\/} (2018), IEEE, pp.~168--179.

\bibitem{yarom2014recovering}
{\sc Yarom, Y., and Benger, N.}
\newblock Recovering openssl ecdsa nonces using the flush+ reload cache
  side-channel attack.
\newblock {\em IACR Cryptol. ePrint Arch. 2014\/} (2014), 140.

\bibitem{yarom2014flush+}
{\sc Yarom, Y., and Falkner, K.}
\newblock Flush+ reload: A high resolution, low noise, l3 cache side-channel
  attack.
\newblock In {\em 23rd $\{$USENIX$\}$ Security Symposium ($\{$USENIX$\}$
  Security 14)\/} (2014), pp.~719--732.

\bibitem{ssh-protocol}
{\sc Ylonen, T., and Lonvick, C.}
\newblock The secure shell (ssh) protocol architecture.
\newblock Tech. rep., 2006.

\bibitem{zhang2011predictive}
{\sc Zhang, D., Askarov, A., and Myers, A.~C.}
\newblock Predictive mitigation of timing channels in interactive systems.
\newblock In {\em Proceedings of the 18th ACM conference on Computer and
  communications security\/} (2011), pp.~563--574.

\bibitem{cross-vm}
{\sc Zhang, Y., Juels, A., Reiter, M.~K., and Ristenpart, T.}
\newblock Cross-vm side channels and their use to extract private keys.
\newblock In {\em Proceedings of the 2012 ACM conference on Computer and
  communications security\/} (2012), pp.~305--316.

\bibitem{zhang2014cross}
{\sc Zhang, Y., Juels, A., Reiter, M.~K., and Ristenpart, T.}
\newblock Cross-tenant side-channel attacks in paas clouds.
\newblock In {\em Proceedings of the 2014 ACM SIGSAC Conference on Computer and
  Communications Security\/} (2014), pp.~990--1003.

\end{thebibliography}

\end{document}